\documentclass[12pt]{iopart}

\usepackage[english]{babel}
\usepackage{hyperref}
\usepackage{cite}
\usepackage{braket}
\usepackage{lineno}

\usepackage[pdftex]{graphicx}
\usepackage{color}

\newcommand{\fig}[1]{Figure~\ref{fig:#1}}
\newcommand{\Fig}[1]{Figure~\ref{fig:#1}}
\newcommand{\sect}[1]{Section~\ref{sec:#1}}
\newcommand{\eq}[1]{Equation~(\ref{eq:#1})}
\newcommand{\lr}[1]{\ensuremath{\left( #1 \right)}}
\renewcommand{\d}{\mathrm{d}}

\newcommand{\bt}{\beta}

\newcommand{\Dl}{\Delta}

\newcommand{\veps}{\varepsilon}

\newcommand{\ta}{\theta}
\newcommand{\mc}{\mathcal}

\newcommand{\av}[1]{\braket{#1}}
\newcommand{\text}[1]{\mathrm{#1}}
\newcommand{\un}[1]{\, \mathrm{#1}}

\begin{document}

\title[Nonequilibrium distribution functions in electron transport]{Nonequilibrium distribution
  functions in electron transport: Decoherence, energy redistribution and dissipation}

\author{Thomas Stegmann$^1$, Orsolya Ujs\'aghy$^2$, Dietrich E. Wolf$^3$}

\address{$^1$ Instituto de Ciencias F\'isicas, Universidad Nacional Aut\'onoma de M\'exico, 62210
  Cuernavaca, Mexico}

\address{$^2$ Department of Theoretical Physics, Budapest University of Technology and Economics,
  H-1521 Budapest, Hungary}

\address{$^3$ Department of Physics, University of Duisburg-Essen and CENIDE, 47048 Duisburg,
  Germany}

\ead{stegmann@icf.unam.mx}

\vspace{5mm}
\begin{indented}
\item \today
\end{indented}

\begin{abstract}
  A new statistical model for the combined effects of decoherence, energy redistribution and
  dissipation on electron transport in large quantum systems is introduced. The essential idea is to
  consider the electron phase information to be lost only at randomly chosen regions with an average
  distance corresponding to the decoherence length. In these regions the electron's energy can be
  unchanged or redistributed within the electron system or dissipated to a heat bath. The different
  types of scattering and the decoherence leave distinct fingerprints in the energy distribution
  functions. They can be interpreted as a mixture of unthermalized and thermalized electrons. In the
  case of weak decoherence, the fraction of thermalized electrons show electrical and thermal
  contact resistances. In the regime of incoherent transport the proposed model is equivalent to a
  Boltzmann equation. The model is applied to experiments with carbon nanotubes. The excellent
  agreement of the model with the experimental data allows to determine the scattering lengths of
  the system.
\end{abstract}

\maketitle

\section{Introduction}
\label{sec:Introduction}

The electron transport in nanosystems shows several quantum phenomena, which makes them promising
for new technological applications. For example, single-walled carbon nanotubes are candidates to
replace silicon in microprocessors. Recently, the first carbon nanotube computer has been realized
\cite{Shulaker2013} as well as the first carbon nanotube transistor which outperforms silicon
\cite{Brady2016}. Topological edge currents, as present for example in quantum Hall insulators, are
considered for new computational devices \cite{Privman1998, Zutic2004}. One fundamental and
important question is how robust the quantum phenomena are against decoherence, energy
redistribution and dissipation. In particular, for efficient carbon-nanotube transistors high
current densities are necessary, where the above mentioned effects become relevant. In recent
experiments, the redistribution of charge carriers in mesoscopic wires \cite{Pothier1997,
  Pothier1997b}, carbon nanotubes \cite{Chen2009, Bronn2013}, quantum Hall edge channels
\cite{LeSueur2010, Altimiras2010, Altimiras2010b} and graphene sheets \cite{Voutilainen2011} has
been investigated. In these experiments it is measured how the energy distribution function $f(E)$
of the charge carriers evolves from a non-equilibrium distribution to an equilibrium Fermi
function. In order to explain the measurements in quantum Hall edge channels, several theories have
been developed \cite{Gutman2009, Levkivskyi2012, Degiovanni2010, Kovrizhin2012, Lunde2010,
  Lunde2016}. Also time-dependent microscopic theories using the density-matrix formalism have been
presented recently \cite{Dolcini2013, Karzig2010, Pepe2012, Dolcini2014}.

We introduce a statistical model, which on the one hand takes into account the effects of
decoherence, energy redistribution and dissipation. This allows to understand the central
observations in the experiments cited above. On the other hand, as the computational demand of our
model is moderate, it can be applied easily to larger systems and may be used as a tool to estimate
the robustness of quantum phenomena.

One succesful approach to take into account the effects of decoherence is B\"uttiker's idea to
introduce virtual reservoirs in the quantum system, where the electrons are absorbed and reinjected
after randomization of their phase and momentum \cite{Buettiker1986}. B\"uttiker's phenomenological
approach can be justified from microscopic theories \cite{Datta1989, Pastawski1991, Datta1991,
  Hershfield1991, Datta1992, Pastawski2001}. To simulate a continuous phase loss in extended
systems, D'Amato and Pastawski generalized this idea by attaching a homogenous distribution of
B\"uttiker probes throughout the system \cite{Amato1990, Cattena2010, Nozaki2012, Nozaki2016}. This
approach has been extended also to a finite voltage and temperature bias \cite{Roy2007}. However,
decoherence may go along with energy redistribution and dissipation which is not considered in these
models.

\begin{figure}[htb]
  \centering
  \includegraphics[scale=0.4]{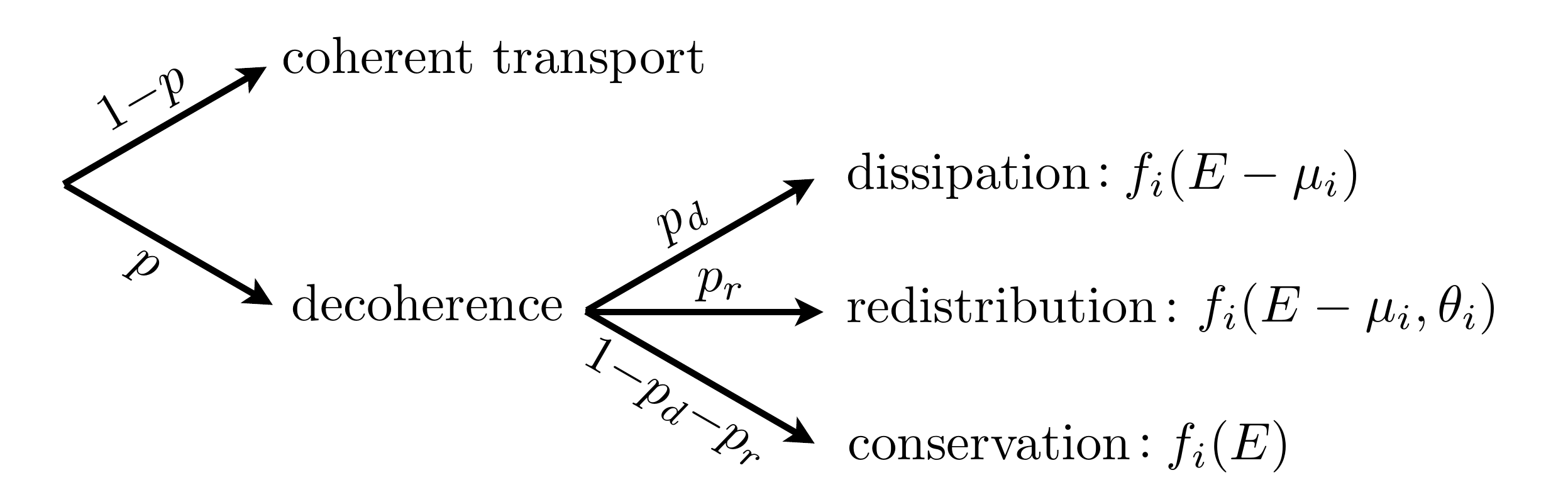}
  \caption{With probability $p$ decoherence regions are introduced, where the phase of the electrons
    is randomized completely. With probability $p_r$ the electrons in a decoherence region
    redistributed their energy within the electron system and relax to a Fermi function
    $f_i(E-\mu_i,\ta_i)$ with chemical potential $\mu_i$ and temperature $\ta_i$. With probability
    $p_d$ the electrons dissipate energy to a heat bath with fixed temperature and are described by
    a Fermi function $f_i(E-\mu_i)$. With probability $1-p_r-p_d$ the electron energy is unchanged
    and they are described by a non-equilibrium distribution function $f_i(E)$.}
  \label{fig:1}
\end{figure}

In our statistical model, we introduce decoherence regions in the quantum system. In these regions
the electron phase is randomized completely, while the transport in between these regions is assumed
to be completely phase coherent. The decoherence regions are introduced with the probability $p$,
see \fig{1}. Hence, the average density of these decoherence regions determines the phase coherence
length $\ell_\phi= 1/p^{1/D}$, where $D$ is the dimension of the system. The decoherence regions can
be localized real defects, as for example in the experiments \cite{Chen2009, Bronn2013}, or virtual
decoherence regions as in B\"uttiker's model. The electrons in the decoherence regions are
characterized by energy distribution functions $f_i$, which describe to which degree
($0 \leq f_i \leq 1$) the density of states at energy $E$ in the $i$th decoherence region is
occupied by electrons. The transport quantity of interest (e.g. the resistance) is averaged over the
ensemble of decoherence configurations, i.e. the ensemble of spatial arrangements of completely
decoherent and completely coherent regions. The ensemble is generated according to a probability
distribution, which may reflect the distribution of real defects in the system. However, in the
following we will consider only random distributions without spatial correlations. In the sense of
the ergodic hypothesis, the ensemble average can be interpreted also as a time average over
fluctuating decoherence configurations. Our statistical model shows the transition from ballistic to
Ohmic conduction under the effect of decoherence \cite{Zilly2009, Stegmann2012}. The model has also
been used to understand several transport experiments on DNA strands \cite{Zilly2010} as well as the
decoherence induced insulator-metal transition in the Anderson model \cite{Zilly2012, Stegmann2014}.

In our previous work, we have assumed that in all decoherence regions the electron energy is
unchanged. In this article, decoherence that goes along with energy redistribution and dissipation
is introduced in the following way: From the set of decoherence regions, which are introduced with
probability $p$, we select a subset with probability $p_r$, see \fig{1}. We assume that in these
regions the electron energy is redistributed within the electron system and relax to a Fermi
function $f_i(E- \mu_i, \ta_i)$ with chemical potential $\mu_i$ and temperature $\ta_i$. In this
way, the effects of electron-electron interaction are introduced, parametrized by the scattering
length $\ell_{\text{red}}= 1/\lr{p p_r}^{1/D}$. In another subset of the decoherence regions, which
is introduced with probability $p_d$, we assume that the electrons dissipate energy to a heat bath
at temperature $\ta_{\text{bath}}$ and are described by a Fermi function $f_i(E-\mu_i)$. In this
way, the effects of electron-phonon interaction (or other types of energy dissipating scatttering)
are introduced with the corresponding scattering length $\ell_{\text{dis}}= 1/\lr{pp_d}^{1/D}$. With
probability $1-p_r-p_d$ the electron energy is unchanged and they are described by a non-equilibrium
distribution function $f_i(E)$. By means of the three parameters
$\{\ell_\phi, \ell_{\text{red}}, \ell_{\text{dis}}\}$, or equivalently, by $\{p,p_r,p_d\}$ we can
tune statistically the degree of decoherence in the quantum system as well as the strength of energy
redistribution and energy dissipation. Note that the arguments of the energy distribution functions
$f_i$ are used to distinguish between the three different cases.

The paper is organized as follows. In \sect{Application}, we present the details of the model and
apply it to a one-dimensional tight-binding chain ($D=1$). In \sect{Properties}, we discuss the
physical properties of the model and demonstrate in \sect{Appl} that it can be used to understand
experiments with carbon nanotubes. The paper is concluded in \sect{Conclusions}.

\section{Statistical model and its application to a tight-binding chain}
\label{sec:Application}

\begin{figure}[htb]
  \includegraphics[scale=0.4]{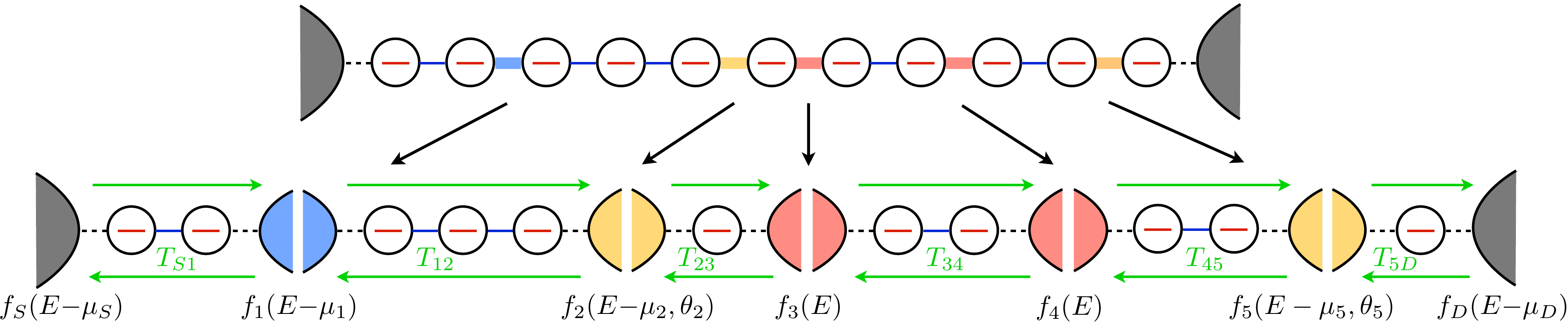}`
  \caption{A tight-binding chain is connected to source and drain reservoirs and driven to
    non-equilibrium by different chemical potentials $\mu_{S/D}$ and temperatures
    $\ta_{S/D}$. Decoherence, energy redistribution and dissipation are introduced in the chain by
    replacing bonds with decoherence reservoirs according to the probabilities in \fig{1}. In the
    blue shaded reservoir of $f_1(E-\mu_1)$ the electrons dissipate energy to a heat bath of fixed
    temperature $\ta_{\text{bath}}$. In the yellow shaded reservoirs of $f_2(E-\mu_2,\ta_2)$ and
    $f_5(E-\mu_5,\ta_5)$ the electrons redistribute their energy. In the red shaded reservoirs of
    $f_3(E)$ and $f_4(E)$ the electron energy is unchanged. Note that the arguments of the
    distribution functions $f_i$ are used to distinguish between the three different cases.}
\label{fig:2}
\end{figure}

We discuss the details of our statistical model by applying it to a tight-binding chain, see \fig{2}
(upper part), described by the Hamiltonian
\begin{equation}
  \label{eq:1}
  H= \sum_{n=1}^{N+1} \veps_n \ket{n}\bra{n} + \sum_{n=1}^N t_{n,n+1}\ket{n}\bra{n+1} + \text{H.c.},
\end{equation}
where the onsite-energies of the $N+1$ sites are denoted by $\veps_n$ and the coupling between the
neighboring sites by $t_{n,n+1}$. The chain is connected at the left and right end to source and
drain reservoirs, which are characterized by Fermi distributions with chemical potentials $\mu_{S/D}$
and absolute temperatures $\ta_{S/D}$. The source and drain reservoirs drive the system into
non-equilibrium due to different chemical potentials and temperatures.

Decoherence regions are introduced by replacing with the probabilities given in \fig{1} the $N$ bonds
of the chain by decoherence reservoirs, see \fig{2} (lower part). For energy conserving decoherence
($p_r=p_d=0)$, we have already used this approach successfully in our previous work
\cite{Stegmann2012, Zilly2012, Zilly2010, Stegmann2013, Stegmann2014}. As the phase coherence is
lost completely in the decoherence regions, the chain is divided into smaller coherent
subsystems. The phase coherent transport within the subsystems is characterized by transmission
functions $T_{i,i+1}(E)$, which can be calculated by means of the non-equilibrium Green's function
(NEGF) method, as we did in our previous work. However, in this article we focus on the effects of
decoherence, energy redistribution and dissipation and assume $T_{i,i+1}(E)= T(E)= 1$ within all
subsystems for all considered energies.

The subdivision into smaller coherent subsystems keeps the computational demand of our model
moderate: The coherent transmission has to be calculated only between neighboring
reservoirs. Moreover each of these calculations is less heavy (for $p>1/N$), as it involves
one-electron Hilbert spaces of average dimension $1/p$ instead of $N$. The ensemble average can be
truncated usually after a few hundred realizations. This makes our model in large systems with weak
decoherence computationally more efficient than the original Pastawski model
\cite{Amato1990}. However, we would like to mention that other work has also been done to improve
the performance of Pastawski's model \cite{Pastawski1991, Nozaki2016}.

The decoherence reservoirs are coupled by rate equations, which express conservation laws depending
on the type of decoherence. In the following two subsections we discuss these rate equations for
general $T$ and solve them analytically for $T=1$.

\subsection{Rate equations for the distribution functions}
\label{sec:Reqs}

The difference between the incoming and outgoing current of electrons with energy $E$ at a
decoherence reservoir $i$ is given by
\begin{equation}
  \label{eq:rate-el}
  \mc{I}(E) \equiv T_{i-1,i}(E) \lr{f_{i-1}{-}f_i} +T_{i+1,i}(E) \lr{f_{i+1}{-}f_i}.
\end{equation}
If this decoherence reservoir is energy conserving, the equation
\begin{equation}
  \label{eq:rate-el2}
  \mc{I}(E) = 0
\end{equation}
has to be fulfilled in steady state. In this case $f_i$ is a non-equilibrium distribution function
$f_i(E)$, while the neighboring reservoirs $f_{i+1}$ and $f_{i-1}$ can be non-equilibrium functions
or Fermi functions.

At an energy dissipating decoherence reservoir electrons loose energy to a heat bath (in the present
case kept at zero temperature, $\ta_{\text{bath}}=0$). Therefore only the charge integrated over all
energies is conserved, i.e. \eq{rate-el} is replaced by
\begin{equation}
  \label{eq:rate-in}
  \int \d E \, \mc{I}(E)=0
\end{equation}
This equation determines the local chemical potential $\mu_i$ for the Fermi distribution
$f_i=f_i(E-\mu_i)$ with absolute temperature equal to $\ta_{\text{bath}}=0$. 

At a decoherence reservoir, where the electron energies are redistributed, \eq{rate-in} must hold,
too. As the energy remains in the electron system, furthermore the equation
\begin{equation}
  \label{eq:rate-inE}
  \int \d E \, E \,\mc{I}(E)=0,
\end{equation}
must be obeyed as well. The two Equations~(\ref{eq:rate-in}) and (\ref{eq:rate-inE}) then determine
the local chemical potential $\mu_i$ and the local temperature $\ta_i$ of the Fermi function
$f_i = f_i(E-\mu_i,\ta_i)$. Obviously, they depend on the distribution functions at the neighboring
decoherence reservoirs.

\subsection{Solution of the rate equations for $T(E)=1$}
\label{sec:Reqs-sol}

For transmission $T(E)=1$ the rate equations can be solved analytically by the Sommerfeld
expansion. Let us consider a given decoherence configuration of $N_D$ decoherence reservoirs of
which $N_D^{\text{ex}}$ are energy exchanging (redistributing and dissipating). Let $N_i$ denote the
number of energy conserving decoherence reservoirs between the $i$th and $(i+1)$th energy exchanging
reservoir. For example, the decoherence configuration in \fig{2} consists of $N_D=5$ decoherence
reservoirs of which $N_D^{\text{ex}}=3$ are energy exchanging. The decoherence configuration
contains only one line of two energy conserving reservoirs and hence $N_2=2.$ Solving \eq{rate-el},
the energy distribution function of the $q_i$th energy conserving reservoir after the $i$th energy
exchanging reservoir is given by
\begin{equation}
  \label{eq:rate-el-sol}
  f_{q_i}=\frac{N_i+1-q_i}{N_i+1} f_i+\frac{q_i}{N_i+1} f_{i+1},
  \quad q_i = 1,2 \dots N_i,
\end{equation}
where $f_{i}$ and $f_{i+1}$ are the Fermi distributions of the confining energy exchanging
decoherence reservoirs (e.g. $f_2$ and $f_5$ in \fig{2}). Hence, the energy distribution functions
of the energy conserving decoherence reservoirs are (thermally broadened) double step functions. The
positions of the steps are determined by the chemical potential of the confining energy exchanging
reservoirs. The step height is given by the position $q_i$ in the line of energy conserving
reservoirs.

Inserting \eq{rate-el-sol} in \eq{rate-in} and using the Sommerfeld expansion for integrals of the
Fermi function \cite{Solyom2009}, we obtain for the chemical potential of the
$i=1\dots N_D^{\text{ex}}$ energy exchanging decoherence reservoirs
\begin{equation}
  \label{eq:mu} 
  \mu_i=\mu_S-\frac{r_i}{N_D+1}(\mu_S-\mu_D).
\end{equation} 
where $r_i=\sum_{j=0}^{i-1} (N_j+1)$ gives the position of the $i$th energy exchanging reservoir in
the line of all decoherence reservoirs. The chemical potential decreases linearly in the line of all
decoherence reservoirs but this is not necessarily the case in real space, as the positions of the
decoherence reservoirs are chosen randomly.

Let $M_i$ be the number of decoherence reservoirs between the $i$th and $(i+1)$th decoherence
reservoir with fixed temperature (e.g. $M_1=4$ in \fig{2}). Using again the Sommerfeld expansion as
well as the previous results, we obtain for the temperatures of the subset of the $s_i$ decoherence
reservoirs after the $i$th reservoir of fixed temperature (e.g.  $s_1 \in \{1,4\}$ in \fig{2})
\begin{equation}
  \label{eq:ta}
  \ta_{s_i}^2=\frac{3}{\pi^2}(\mu_i{-}\mu_{i+1})^2\frac{s_i}{M_i{+}1}
  \lr{1{-}\frac{s_i}{M_i{+}1}} +\left(\ta_i^2{-}\ta_{i+1}^2\right)\frac{s_i}{M_i{+}1} +\ta_i^2.
\end{equation}
Note that the second term in \eq{ta} vanishes, if all dissipating reservoirs have the same
temperature. The third term vanishes, if this temperature is set to zero.

Starting from the Boltzmann equation and assuming a linear decay of the chemical potential along the
chain, it is discussed in Refs.~\cite{Nagaev1995, Kozub1995,Naveh1998,Huard2007} that the
temperature profile along the chain is determined by the differential equation
\begin{equation}
  \label{eq:taBoltzmann}
  \frac{\pi^2}{6} \frac{d^2 \ta^2}{dx^2}= -(\mu_S-\mu_D)^2 +\bt^2\lr{\ta^5-\ta_\text{bath}^5}
\end{equation}
with the boundary conditions $\ta(0)=\ta(N+1)=\ta_{\text{bath}}$. The first term in \eq{taBoltzmann}
denotes the electron-electron scattering, whereas the second term controls via the parameter $\beta$
the strength of the electron-phonon scattering. In \sect{Properties} we will compare solutions of
\eq{taBoltzmann} with our model.

\subsection{Ensemble averaged distribution functions}
\label{sec:avgs}

Using the analytical solution of the preceding section, we can calculate for a given decoherence
configuration the energy distribution functions at those bonds, which have been replaced by
decoherence reservoirs. Those bonds, which have not been replaced and hence are part of coherent
subsystems, are described by a more complicated density matrix. However, we do not consider only a
single fixed decoherence configuration but an ensemble of random realizations. In order to determine
the ensemble averaged energy distribution function $\av{f}$ for a given bond $x$ of the chain, we
consider from the ensemble of decoherence configurations only those realization where the bond $x$
has been chosen as a decoherence bond. The ensemble averaged distribution function $\av{f}$ is then
given by the arithmetic average over this subset.

\section{Properties of the model}
\label{sec:Properties}

We discuss the properties of the presented model by means of a tight-binding chain of $N+1=101$
sites (i.e. 100 bonds) with transmission $T(E)=1$. The chemical potentials of the source and drain
reservoirs are used to define the energy scale and hence, set to $\mu_S= -\mu_D = 1$. The
temperature of the source and drain, as well as the temperature of the dissipating decoherence
reservoirs is set to $\ta_{\text{bath}}=0$. We consider ensembles of $2 \cdot 10^4 /p$
configurations.


\begin{figure}[t]
\centering
\includegraphics[scale=0.5]{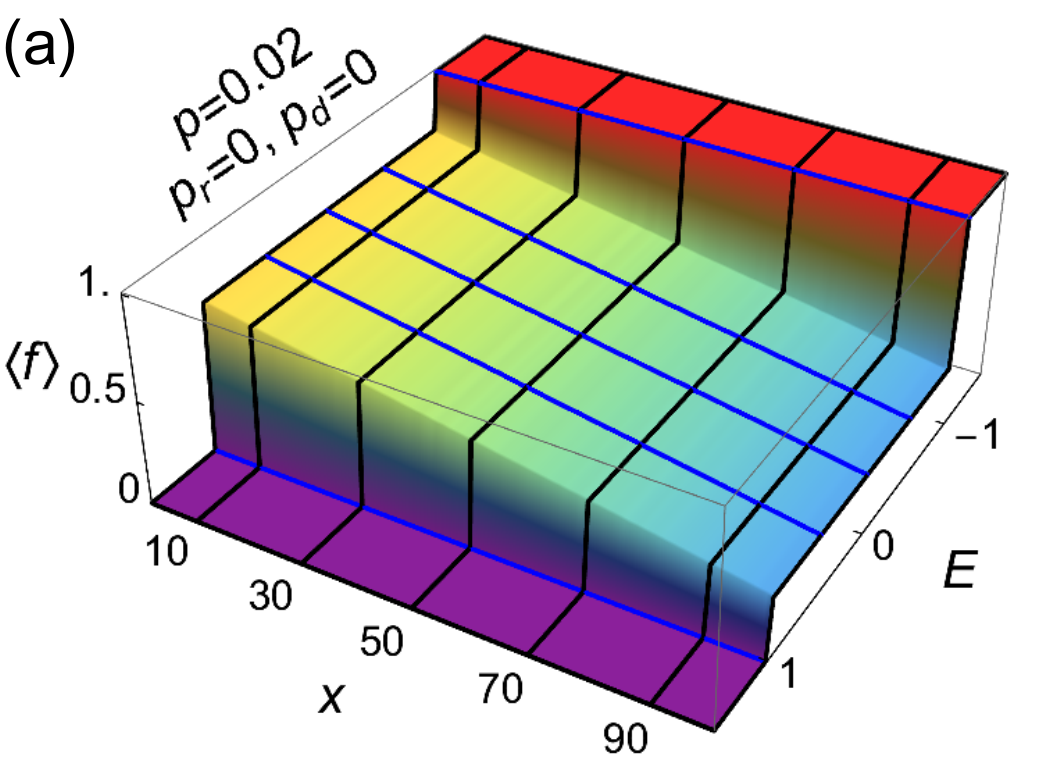}
\includegraphics[scale=0.5]{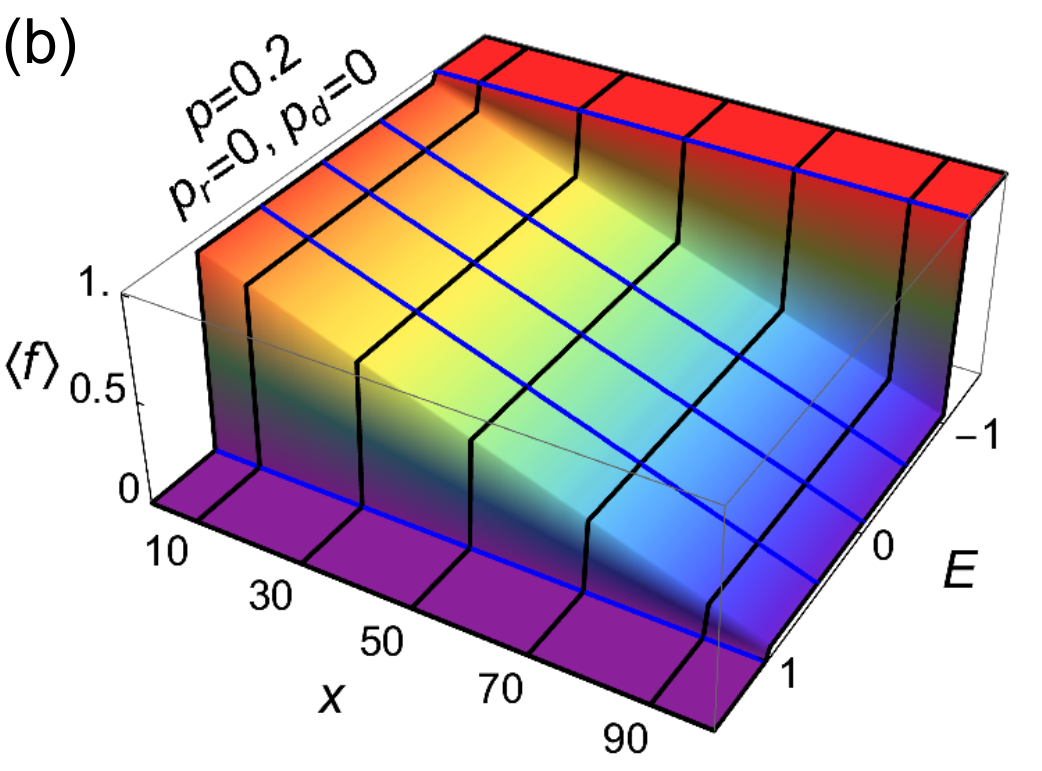}
\caption{Ensemble averaged energy distribution function $\av{f}$ of a tight-binding chain of length
  $N=100$. The decoherence probabilities have been fixed to $p=0.02$ (a) and $p=0.2$ (b) without
  energy redistribution $p_r=0$ nor energy dissipation $p_d=0$. Fixing the position $x$ in the
  chain, $\av{f}$ shows two jumps at the chemical potential of source and drain $\mu_{S/D}= \pm
  1$. At a fixed energy $\mu_D < E < \mu_S$, the distribution function is changing linearly and at
  the chain ends it jumps to $\av{f}=1$ of the source and $\av{f}=0$ of the drain.}
\label{fig:3}
\end{figure}

\Fig{3} shows the ensemble averaged energy distribution function $\av{f}$ as a function of the
electron energy $E$ and the position $x$ in the chain for two degrees of decoherence, $p=0.02$ (a)
and $p=0.2$ (b). Neither energy redistribution ($p_r=0$) nor energy dissipation ($p_d=0$) are
present. At a fixed position $0<x<101$, the distribution function $\av{f}$ shows two steps at
$E=\mu_{S/D}$ and is constant otherwise. A constant $0 \leq \av{f} \leq 1$ is observed in the energy
range $\mu_S < E < \mu_D$, because the source is tending to occupy states in the chain while the
drain is tending to empty these states. As the energy of the electrons is unchanged, the
distribution function is constant in this energy range. Furthermore, for $E< \mu_D$ all states are
occupied and for $E> \mu_S$ all states are unoccupied. At a fixed energy $\mu_D < E < \mu_S$, the
distribution function decreases linearly along the chain. At the chain ends the distribution
function shows two jumps to $\av{f}=1$ at the source and $\av{f}=0$ at the drain [\footnote{Note
  that the energy distribution functions $f(E-\mu_{S/D})$ of source and drain are not shown in
  \fig{3}.}]. The height of these jumps decreases if the degree of decoherence increases. In the
following, the effects of energy redistribution and dissipation are discussed focusing on the cases
of strong decoherence ($p=0.2$) and weak decoherence ($p=0.02$).


\begin{figure}[t]
\centering
\includegraphics[scale=0.36]{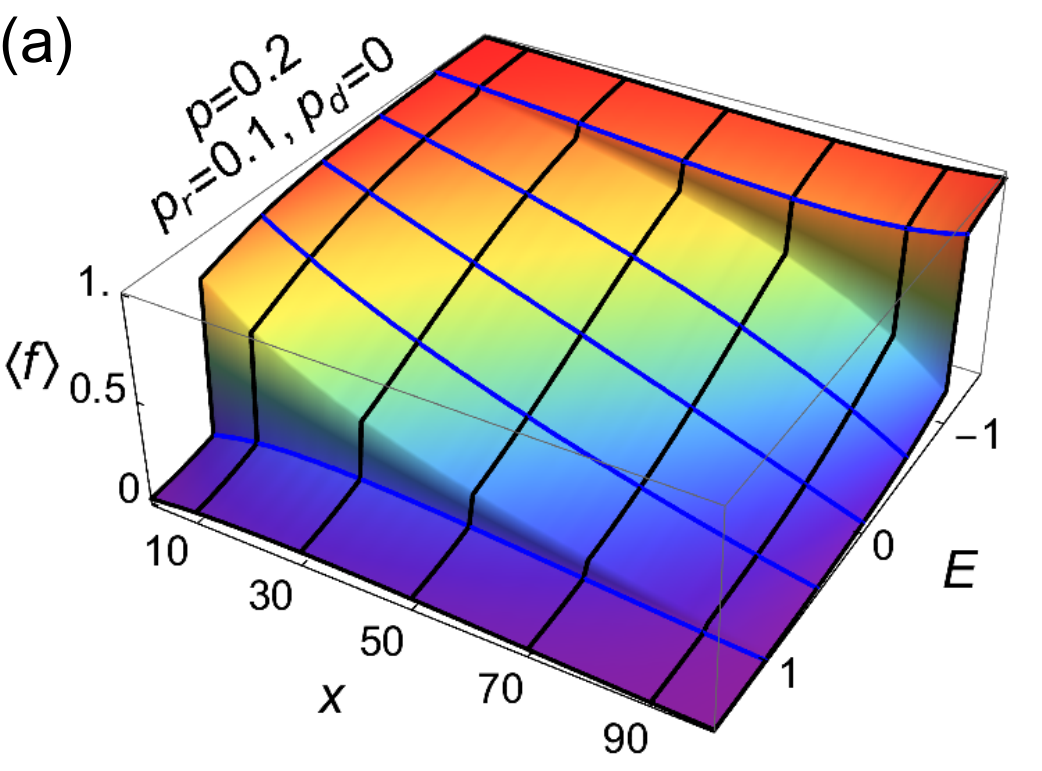}
\includegraphics[scale=0.36]{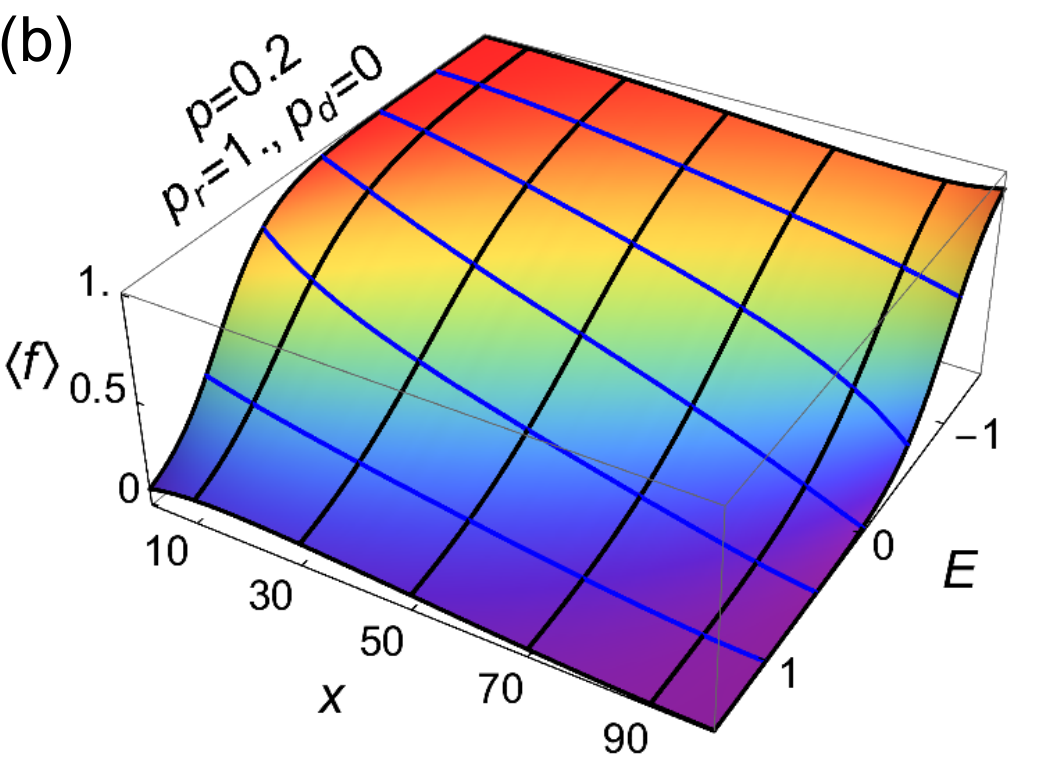}
\includegraphics[scale=0.36]{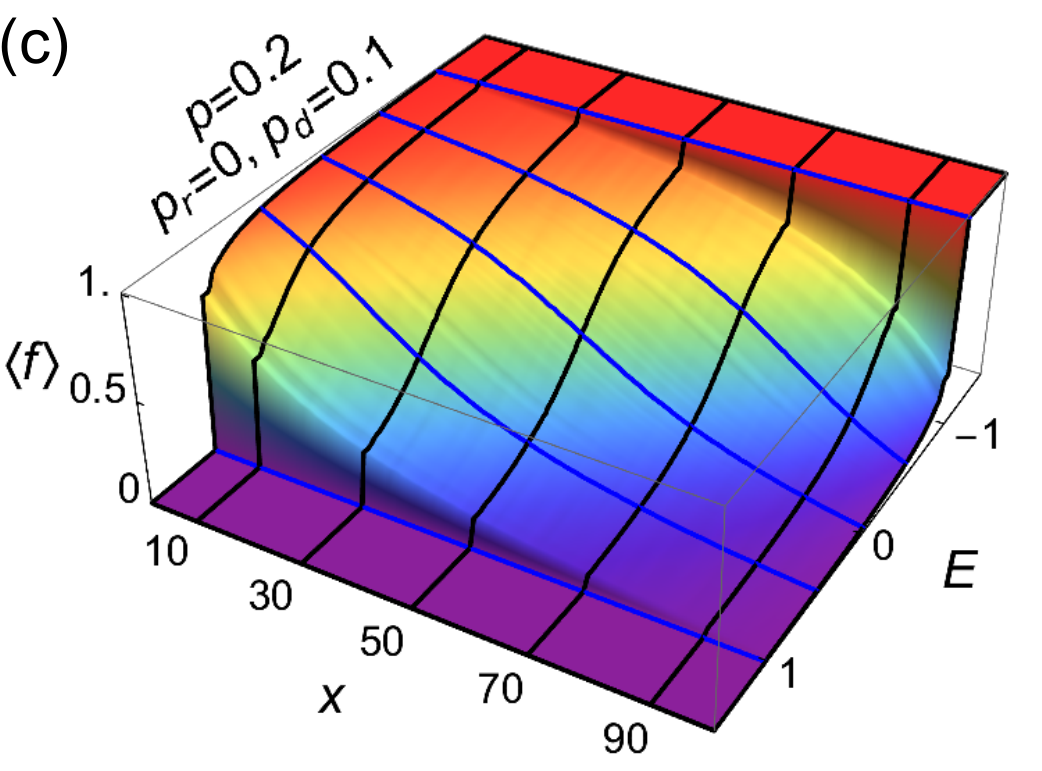}
\includegraphics[scale=0.36]{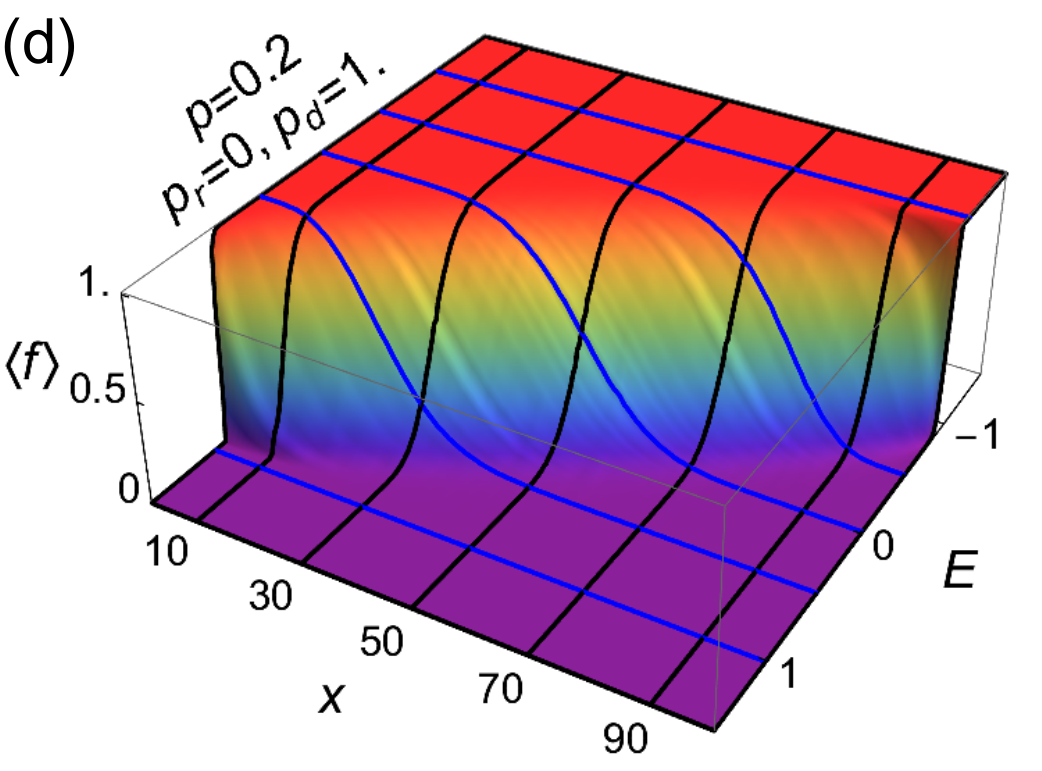}\\[3mm]
\includegraphics[scale=0.36]{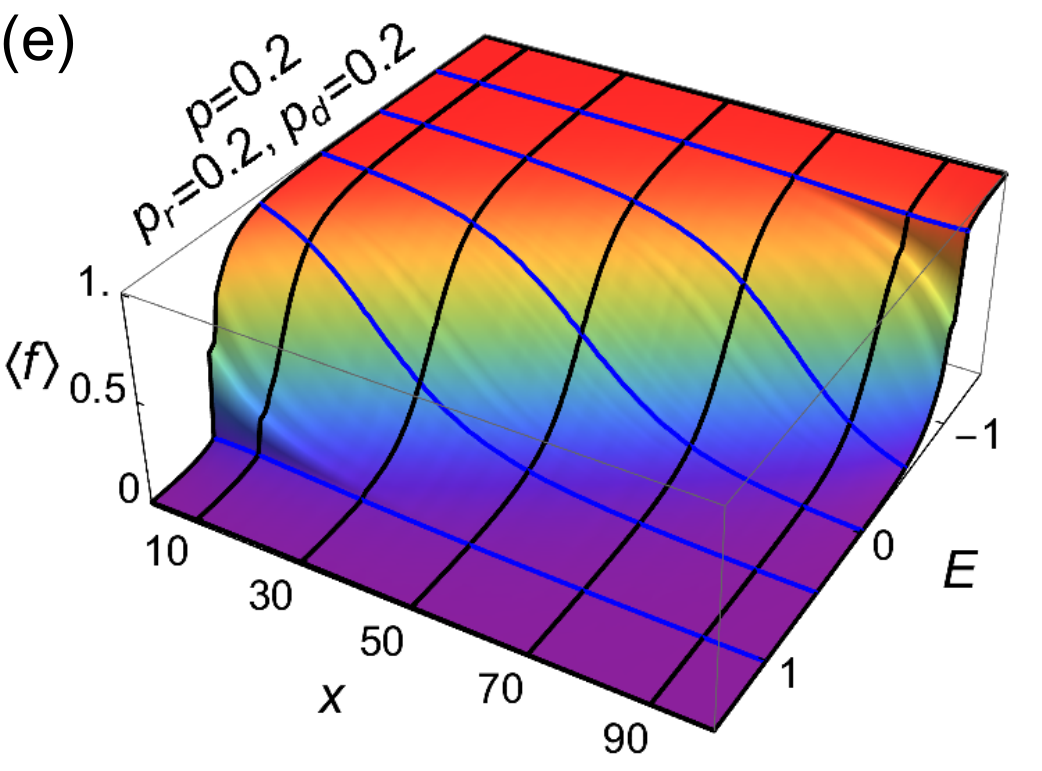}
\includegraphics[scale=0.36]{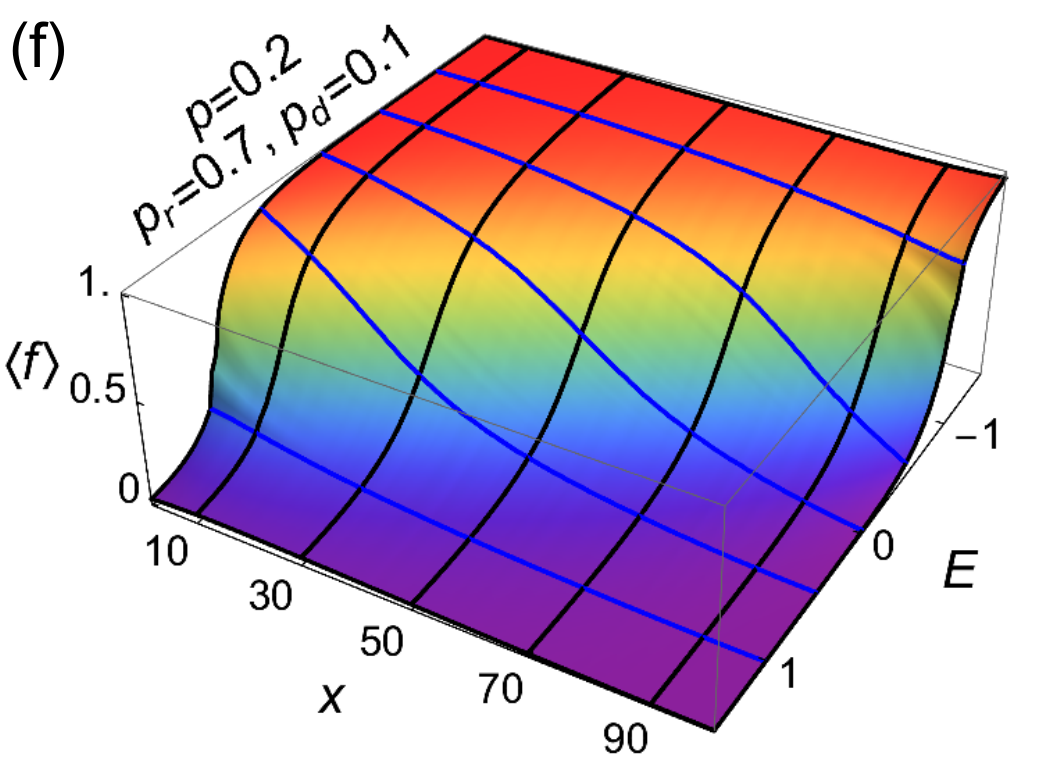}
\includegraphics[scale=0.36]{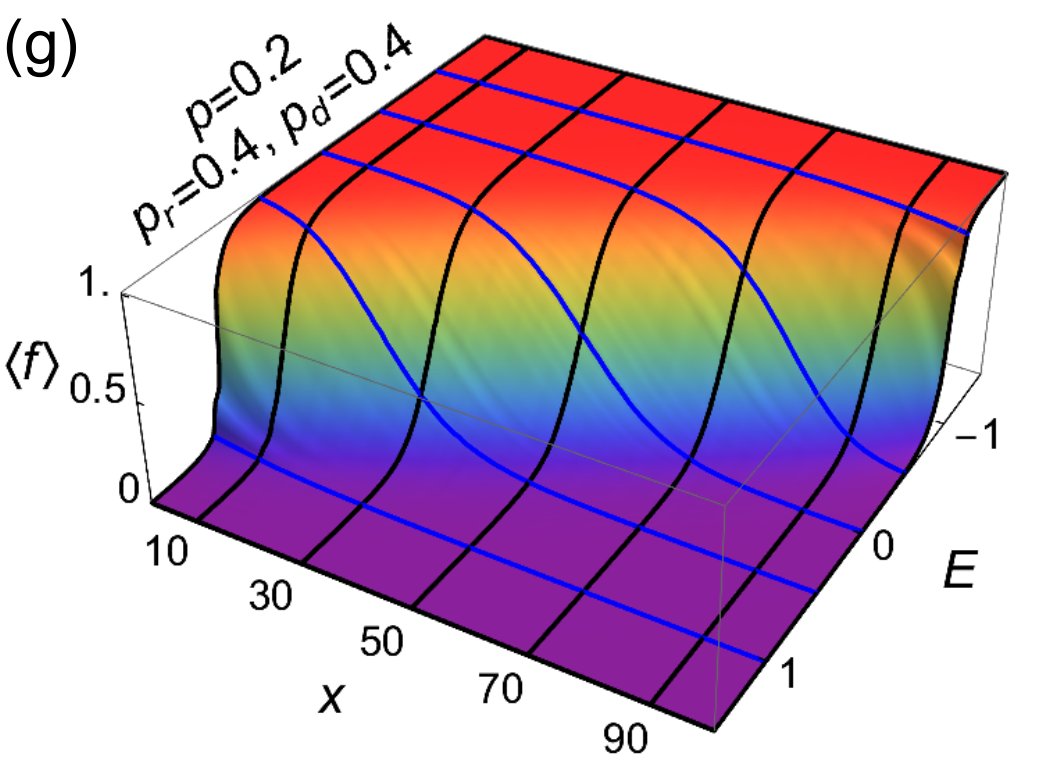}
\includegraphics[scale=0.36]{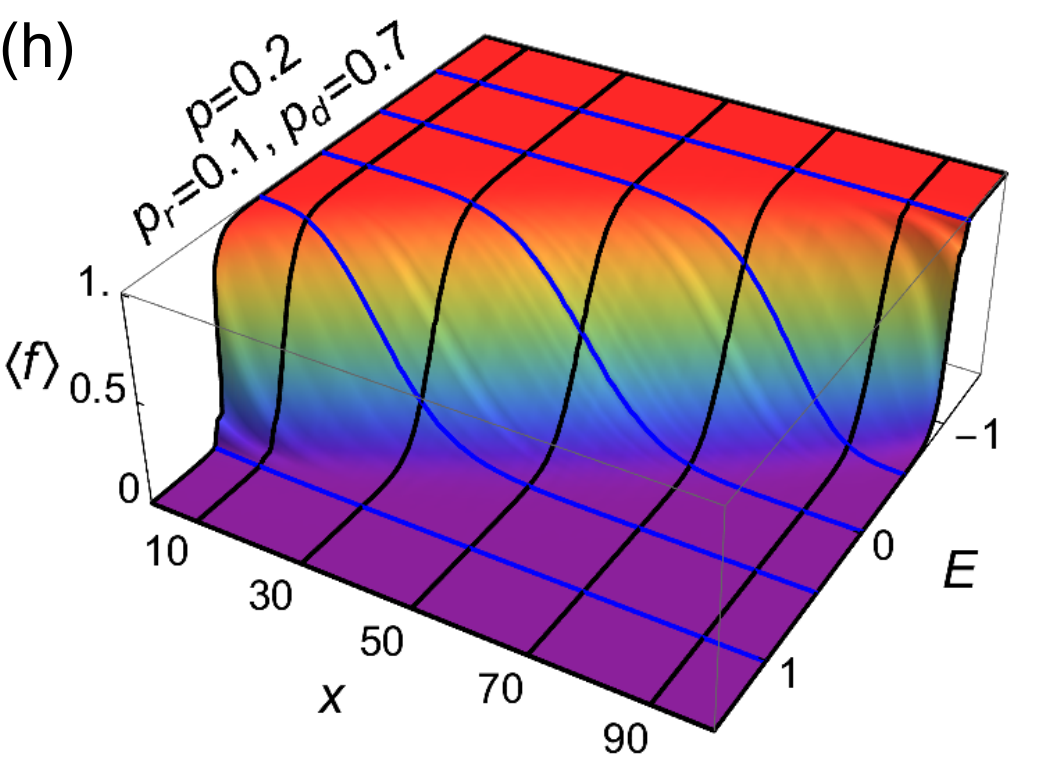}
\caption{Energy distribution functions $\av{f}$ under the effect of energy redistributing decoherence and
  energy dissipating decoherence. The degree of decoherence is fixed to $p=0.2$. In the first row,
  either energy redistribution (a,b) or energy dissipation are present (c,d). In (a,c) $p_r$ and $p_d$
  are weak, while in (b,d) they are strong. In the second row, both types of scattering are
  mixed. We observe that energy redistribution transforms the double step function into a very smooth
  Fermi function, whereas energy dissipation causes a very steep Fermi function.}
\label{fig:4}
\end{figure}

\Fig{4} shows the energy distribution function for $p=0.2$ in presence of energy redistribution and
dissipation. In the first row, either $p_r$ (a,b) or $p_d$ (c,d) are nonzero. For low $p_{r/d}$
(a,c) the jumps in the distribution functions are smoothed but still visible. In the range
$\mu_D < E < \mu_S$ the distribution function is no longer constant but arched. In the case of high
$p_{r/d}$ (b,d), the distribution function does not show a double step at the chemical potential of
source and drain, anymore. Instead, it can be described by a Fermi function as will be discussed
below. The distribution functions are quite different: For strong energy redistribution (b), the
distribution function is much broader because the total energy of the electron system is
unchanged. For strong energy dissipation (d), the distribution function is very steep, because the
electrons dissipate their excess energy to a heat bath at zero temperature. Note that the
redistribution of energy present in (a,b) allows for $\av{f(E>\mu_S)}>0 $ and $\av{f(E<\mu_D)}<1$
corresponding to Auger processes. This is not possible if energy is dissipated to heat baths with
$\ta_\text{bath}=0$, see (c,d). In the second row of \fig{4}, energy redistribution and dissipation
are both present. Although the energy distribution functions are quite similar for some parameters,
we can detect by means of their overall shape, if energy redistribution or energy dissipation are
present and, if one of these processes dominates.

\begin{figure}[htb]
\centering
\includegraphics[scale=0.72]{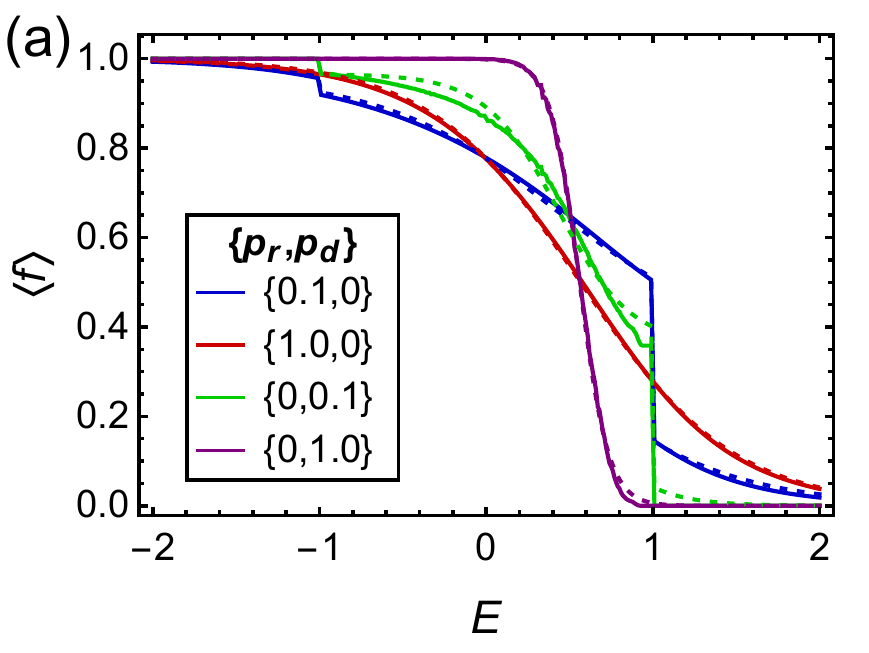}
\includegraphics[scale=0.72]{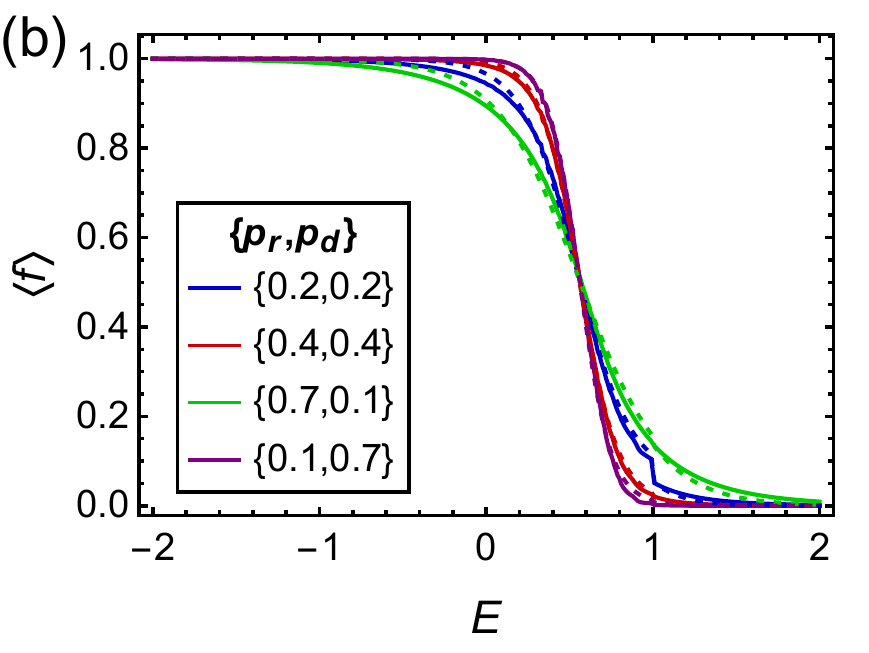}
\caption{Ensemble averaged distribution function (solid curves) for $p=0.2 $ and $x=20$. Fits with
  the model in \eq{model} (dashed curves) to the numerical data match excellently for all studied
  parameters.}
\label{fig:5}
\end{figure}


The ensemble averaged energy distribution functions $\av{f}$ give plenty of information about the
electronic system of the tight-binding chain but they are difficult to analyze. Hence, our aim is to
extract from $\av{f}$ some few meaningful physical parameters, which characterize the system. This
can be done by means of the model
\begin{eqnarray}
  \label{eq:model}
  \hspace*{-10mm}
  \av{f}_{\text{fit}}(E,x)= a_S(x) f(E{-}\mu_S) +a_D(x) f(E{-}\mu_D) +a_{rd}(x) f(E{-}\mu(x),\ta(x)),
\end{eqnarray}
where $a_{S/D} \geq 0$ give the fraction (or relative densities) of electrons which follow the
energy distribution function of source and drain, respectively. The parameter $a_{rd} \geq 0$ gives
the fraction of thermalized electrons, which are described by a Fermi function with chemical
potential $\mu$ and temperature $\ta$. The five parameters $\{a_{S/D/rd}, \ta, \mu \}$ are
determined from the numerical data in the following way. The two parameters $a_{S/D}$ are calculated
from the height of the steps at $\mu_{S/D}$ in the distribution functions, because source and drain
are at zero temperature. The remaining parameter $a_{rd}$ can be determined by using
$f(E \to -\infty)= a_S+a_D+a_{rd}= 1$. Hence, not five but only two parameters $\ta$ and $\mu$ are
determined by a fit to the numerical data. Due to the symmetry of the boundary conditions and the
homogeneity of the chain, the fit parameters fulfill the symmetry relations $a_S(x)= a_D(N-x)$,
$a_{rd}(x)= a_{rd}(N-x)$, $\mu(x)= -\mu(N-x)$ and $\ta(x)= \ta(N-x)$. In general the model
\eq{model} describes extremely well the numerical data. A typical example is shown in \fig{5}.


\begin{figure}[htb]
\centering
\includegraphics[scale=0.7]{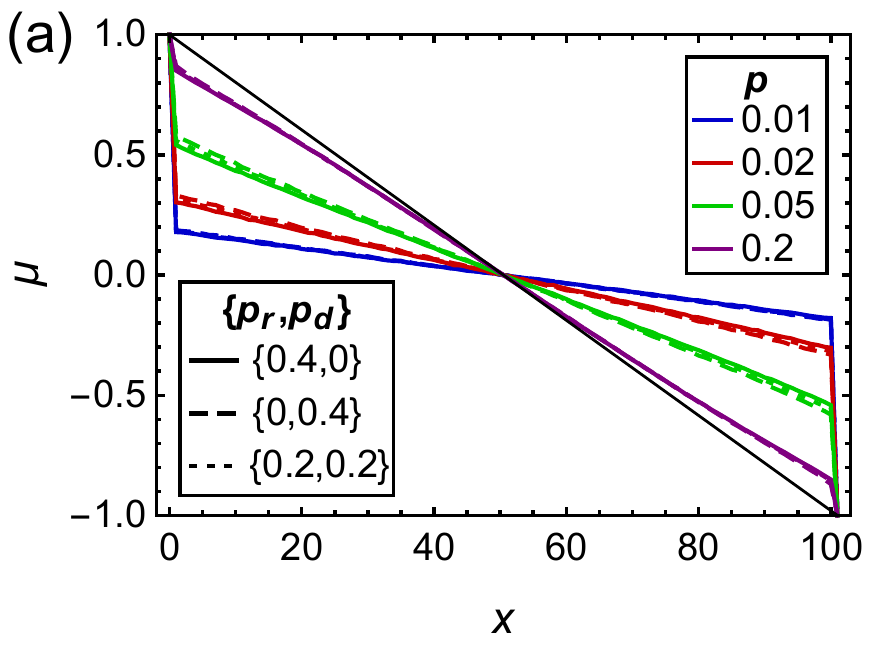}
\includegraphics[scale=0.7]{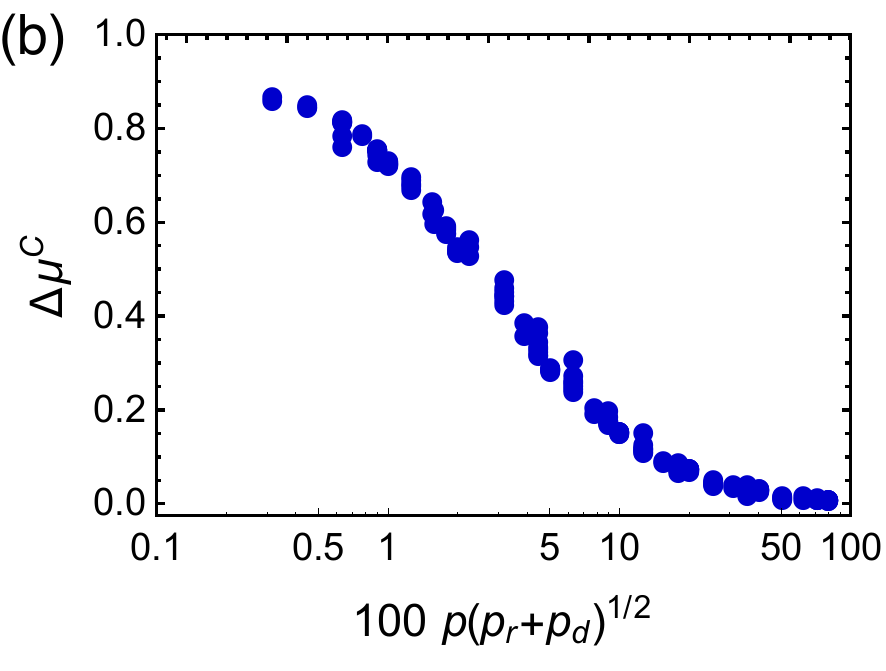}
\caption{Left: Chemical potential in the chain. Within the chain the chemical potential decays
  linearly, while at the chain ends it jumps to $\mu_{S/D}= \pm 1$. These jumps can be attributed to
  the contact resistance of the chain. The thin black line represents a linear profile from the
  source to the drain. Right: Jumps of the chemical potential at the chain ends, calculated for
  various system parameters, as a function of $p\sqrt{p_r+p_d}$.}
\label{fig:6}
\end{figure}

The chemical potential $\mu(x)$, determined by fitting the numerical data with \eq{model}, is shown
in \fig{6}~(a). Inside the chain the chemical potential decays linearly, while at the chain ends it
jumps abruptly to the fixed chemical potential $\mu_{S/D}=\pm 1$ of source and drain,
respectively. As the chemical potential characterizes the potential energy of the electrons, its
course can be interpreted as the voltage drop along the chain \cite{Datta1997, McLennan1991}. Hence,
we observe an Ohmic linear voltage profile inside the chain while the jumps at the chain ends
$\Dl \mu^C$ can be attributed to the contact resistance of the chain \cite{Lan2008, Choi2008,
  Liu2008}. The jumps are largest for low degree of decoherence $p$. They are symmetric because the
source and drain reservoirs are identically coupled. If $p$ increases, $\Dl \mu^C$ diminishes and a
linear voltage profile from the source to the drain is approached, see the thin black line. Such a
linear voltage profile is one of the initial assumptions of the Boltzmann ansatz leading to
\eq{taBoltzmann}. By contrast, in our model the chemical potential profile is not an assumption but
is calculated. Furthermore, we can model the entire regime from phase coherent to Ohmic
transport. \Fig{6}~(b) shows that $\Dl \mu^C$ does not depend on $p,p_r,p_d$ individually but only
on the combination $p\sqrt{p_r+p_d}$. Other combinations of these three parameters would not lead to
a collapse of the data on a single curve.

Note that only if $a_{rd}=1$ all electrons are described by a Fermi function. In general $\mu$
describes only the fraction of thermalized electrons, $a_{rd}<1$. This regime, in which a voltage
drop in the Ohmic sense cannot be defined, is not covered by the Boltzmann ansatz in
Refs.~\cite{Nagaev1995, Kozub1995,Naveh1998,Huard2007} where all electrons are described by a Fermi
function.


\begin{figure}[htb]
\centering
\includegraphics[scale=0.7]{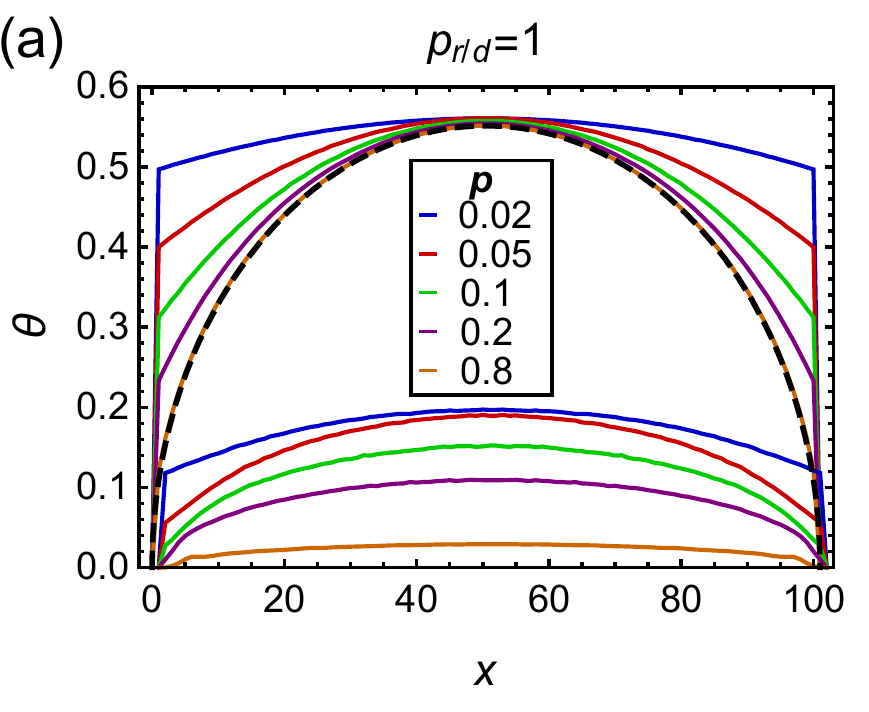}
\includegraphics[scale=0.7]{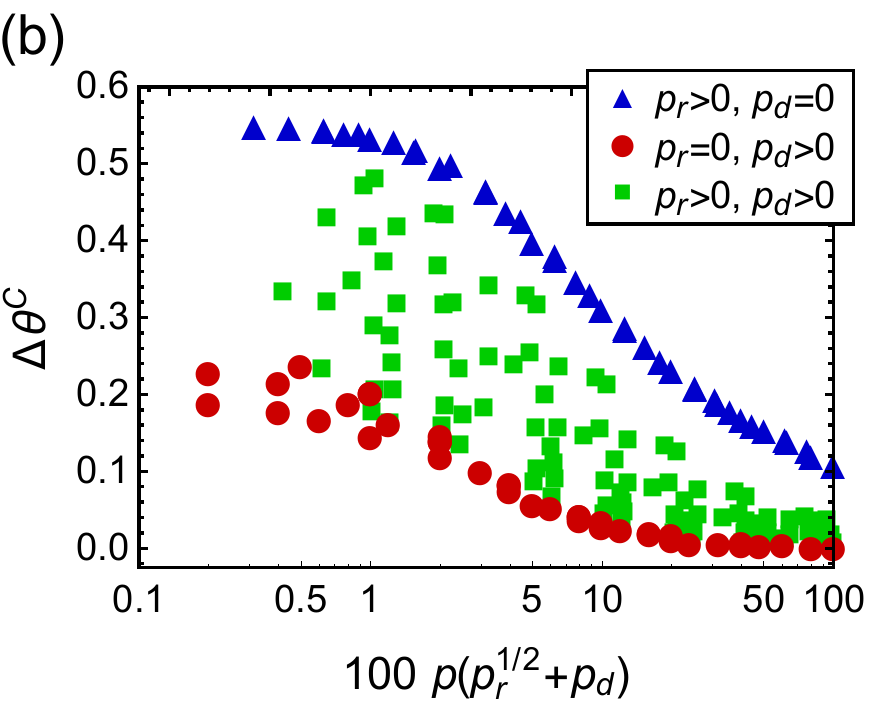}\\[3mm]
\includegraphics[scale=0.7]{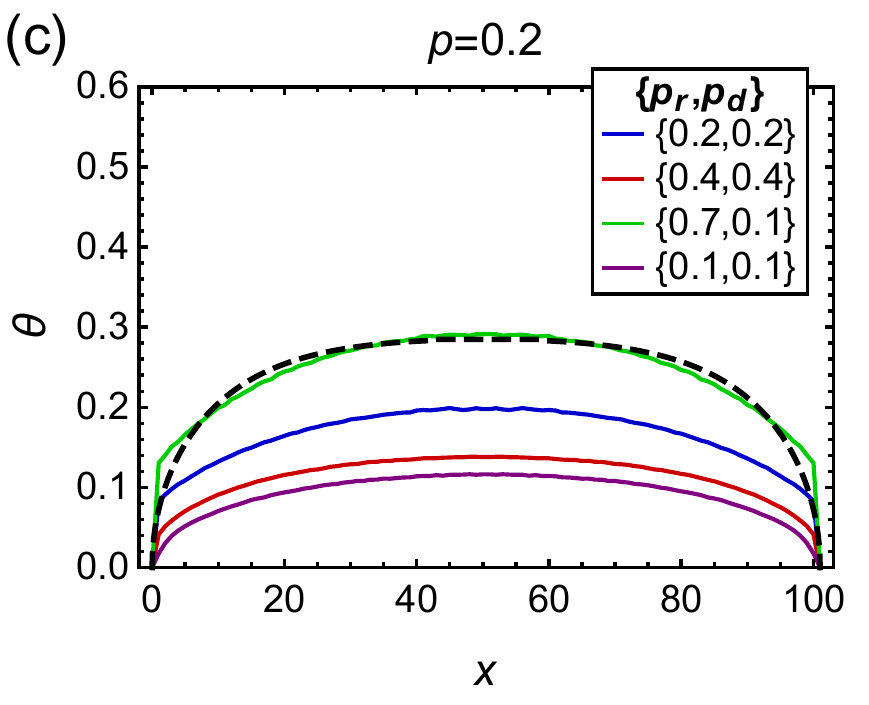}
\includegraphics[scale=0.7]{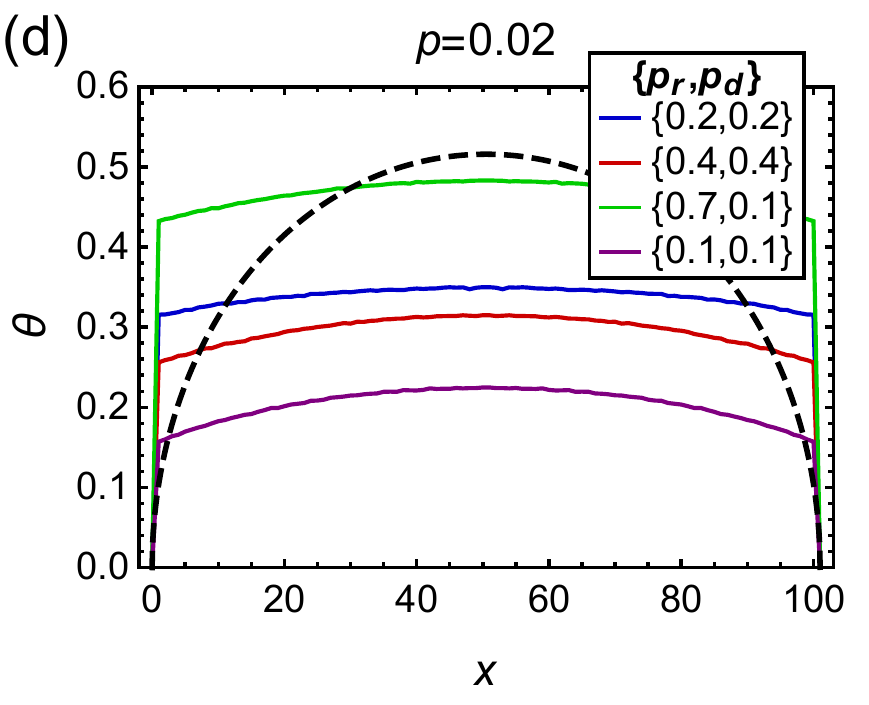}
\caption{(a): Temperatures profiles $\ta(x)$ along the chain for various degree of decoherence $p$
  in the case that either $p_r=1$ (top curves) or $p_d=1$ (bottom curves). When $p$ is increasing
  $\ta$ for $p_r=1$ approaches the black dashed curve, which is the solution of the Boltzmann model
  in \eq{taBoltzmann} for $\bt=0$. (b): Temperature jumps at the chain ends as a function of
  $p(\sqrt{p_r}+p_d)$. (c,d): Temperatures for non-zero $p_r$ and $p_d$. The black dashed curve
  represents the temperatures from the Boltzmann ansatz with $\bt=22$ (c) and $\bt=2.2$ (d). In the
  case of strong decoherence ($p=0.2$) both models agree qualitatively, whereas they disagree in the
  case of weak decoherence ($p=0.02$), because the Boltzmann model does not describe coherent
  transport.}
\label{fig:7}
\end{figure}

The temperature profile along the chain is shown in \fig{7}~(a) for $p_r=1$ (top curves) and $p_d=1$
(bottom curves). When $p$ increases the temperature profile for $p_r=1$ approaches the black-dashed
curve. This curve gives the solution of \eq{taBoltzmann} in the case $\bt=0$, which in turn is
equivalent to the temperature profile in \eq{ta} for $p=p_r=1$. Hence, in this limit case both
models give identical results. When $p$ decreases, the temperature in the center of the chain stays
constant but increases at the chain ends. For $p_r=1$ the Joule heat produced in the system can only
be evacuated through the source and drain kept at zero temperature. For $p<1$ this causes abrupt
jumps of the temperature at the chain ends, similar to those in the chemical potential. These
temperature jumps $\Dl \ta^C$ can be attributed to the thermal contact resistance of the chain
\cite{Fujii2005, Pop2006, Pop2008}. 

The temperature profile for $p_d=1$ approaches zero when $p$ is increased. This property is
consistent with the fact that the electrons dissipate energy to heat baths at zero temperature. It
is obtained also from \eq{taBoltzmann} in the limit $\bt \to \infty$. A finite temperature for $p<1$
arises in our model from statistical fluctuations. Physically this corresponds to a reduction of the
cross-section for electron-phonon scattering, which prevents the electrons from dissipating all
their excess energy to the heat baths.

The temperature jumps $\Dl \ta^C$ are shown in \fig{7}~(b) as a function of $p(\sqrt{p_r}+p_d)$ for
various system parameters. In general, energy redistribution (blue triangles) leads to higher values
of $\Dl \ta^C$ than energy dissipation (red dots). If either $p_r$ or $p_d$ are zero, the data
points collapse onto single curves. If both are non-zero (green squares) the jumps $\Dl \ta^C$ and
hence, the temperature along the chain depend on the individual values of $p_r$ and $p_d$. Note that
the scaling properties of $\Dl \ta^C$ and the $\Dl \mu^C$ are different, compare with
\fig{6}~(b). If the data in \fig{7}~(b) are plotted as a function of $p\sqrt{p_r+p_d}$, a data
collapse on single curves (blue triangles and red dots) is not observed.

When either $p_r$ or $p_d$ are nonzero but less than 1, the temperature profiles show qualitatively
the same behavior as in \fig{7}~(a). However, in this case only the fraction $a_{rd}$ of the
electrons is characterized by the temperature $\ta$. This reflects the fact that the system is in
non-equilibrium and cannot be described by a thermodynamic temperature. \Fig{7}~(c,d) show the
temperature profiles in the case that $p_r$ and $p_d$ have both non-zero values. In the case of
strong decoherence ($p=0.2$) the calculated temperature profile agrees qualitatively with the
temperature profile from the Boltzmann model \eq{taBoltzmann} using the fit parameter $\bt=22$,
compare the solid green curve and the black dashed curve. For low decoherence ($p=0.02$) the black
dashed curve, representing the solution of \eq{taBoltzmann} with $\bt=2.2$, disagrees with the
temperature from our model, which is almost constant inside the chain with steep temperature drops
at the chain ends. This disagreement is not surprising because the Boltzmann ansatz in
\eq{taBoltzmann} does not describe coherent transport.


\begin{figure}[htb]
\centering
\includegraphics[scale=0.7]{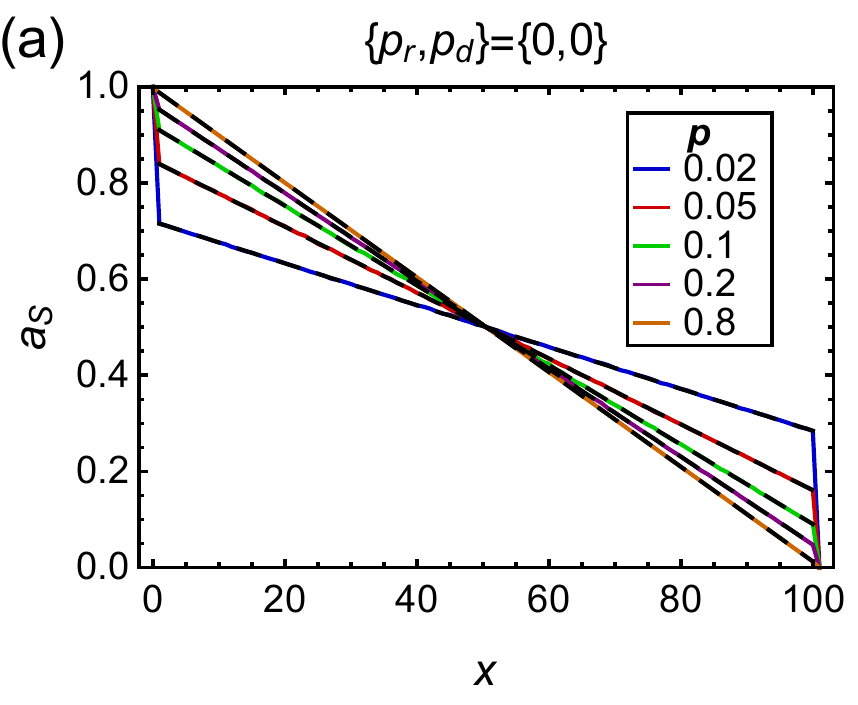}
\includegraphics[scale=0.7]{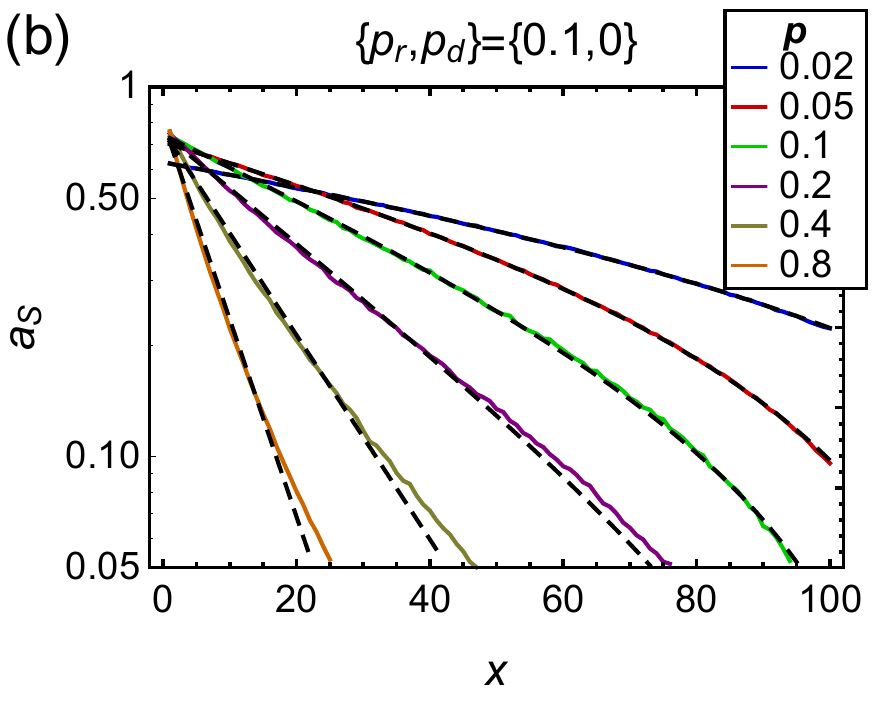}\\
\includegraphics[scale=0.7]{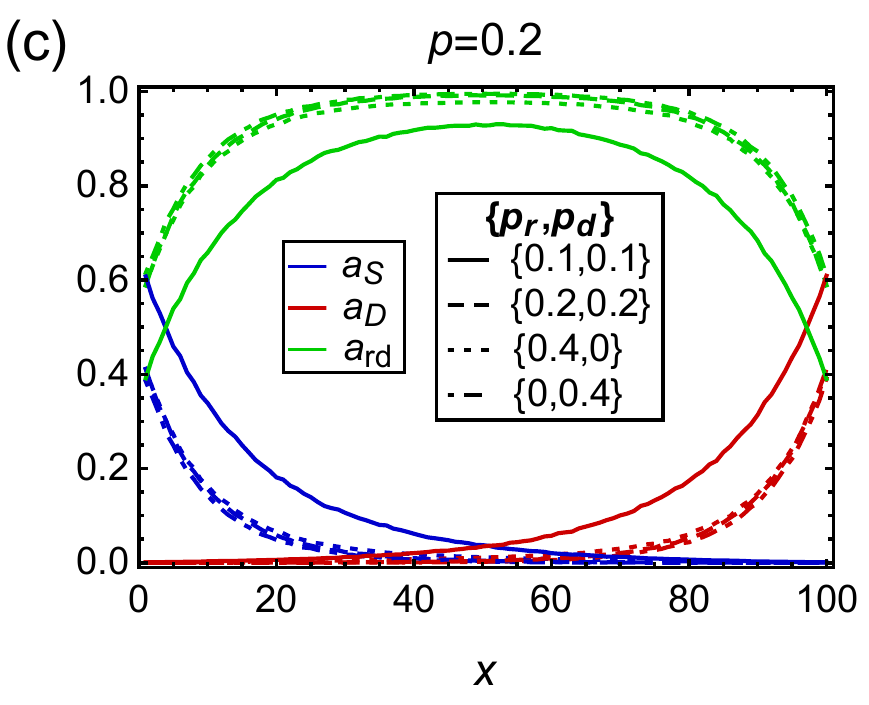}
\includegraphics[scale=0.7]{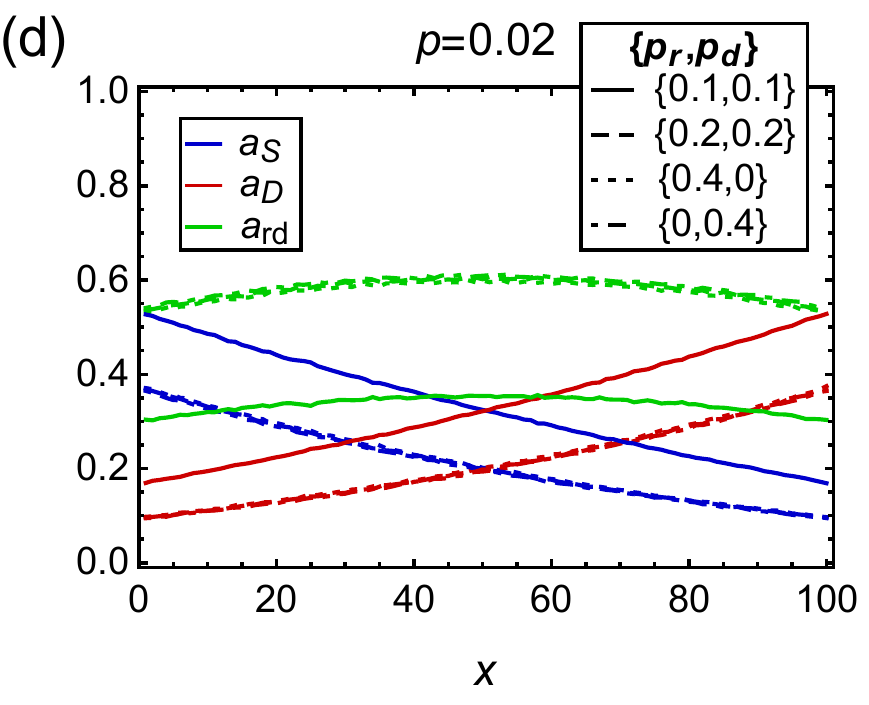}
\caption{Fractions (or relative densities) of the electrons along the chain. $ a_{S/D} $ represent
  the fraction of electrons which are unthermalized, while $a_{rd}$ gives the fraction of
  thermalized electrons. $a_S$ changes linearly if $p_r=p_d=0$, whereas it is dominated by an
  exponential decay if $p_r$ or $p_d$ are nonzero. The dashed black curve in (b) represent fits
  using \eq{asd}.}
\label{fig:8}
\end{figure}

\vspace{7mm}

The fractions (or densities) of electrons, which follow a certain distribution function in
\eq{model}, are shown in \fig{8}. The fraction of electrons following the source can be described by
\begin{equation}
  \label{eq:asd}
  a_S(x) = \frac{1+\tilde{a}}{2}\exp\lr{-\frac{x}{\xi}} + \tilde{a}\frac{x}{N}\exp\lr{-\frac{N}{\xi}}
\end{equation}
with the two parameters $\tilde{a}$ and $\xi$. In the case $p_r+p_d=0$, the characteristic length
$\xi \to \infty$ and $a_S$ decays linearly, see \fig{8}~(a). This behavior can be understood by
taking into account the symmetry relation $a_S(x)= a_D(N-x)$ and the particle conservation
constraint. In the limit case of coherent transport ($p \sim 0$), we obtain $a_S=a_D=0.5$, because
no scattering takes place in the chain and hence, its states are occupied equally by source and
drain.

If $p_r$ or $p_d$ are nonzero, see for example \fig{8} (b), $a_S$ is dominated by the exponential
decay close to the source while farther away linear corrections have to be taken into
account. \Fig{8} (c,d) show that the electron densities are functions of $p_r+p_d$. The fact that
$a_{S/D/rd}$ and $\mu$ depend only on the sum $p_r+p_d$ while only $\ta$ depends on the individual
values of $p_{r/d}$ makes the distribution functions for constant $p_r+p_d$ similar but not
identical.


Due to the fact that we assumed for the transmission in the chain $T(E)=1$, the distribution
functions show the following scaling law. Considering two different sets of parameters
$\{N,p,p_r,p_d\}$ and $\{N',p',p_r',p_d'\}$, the distribution functions of long chains ($N \gg 1$)
are identical, if the parameters fulfill $Npp_{r/d}= N'p'p_{r/d}'$. Note that this scaling law is
restricted by the constraint $0 \leq p,p_{r/d},p_r+p_d \leq 1$. The scaling law breaks down if
$p \sim 0$, because effects of the coherent transport become important.


\begin{figure}[t]
\centering
\includegraphics[scale=0.36]{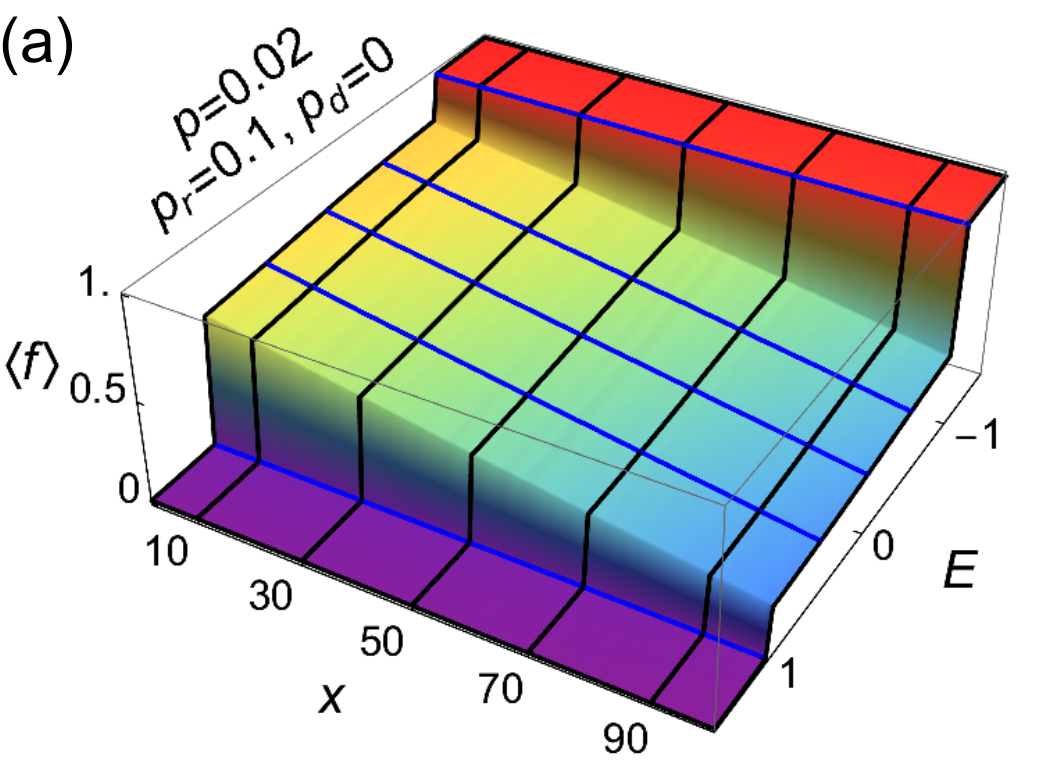}
\includegraphics[scale=0.36]{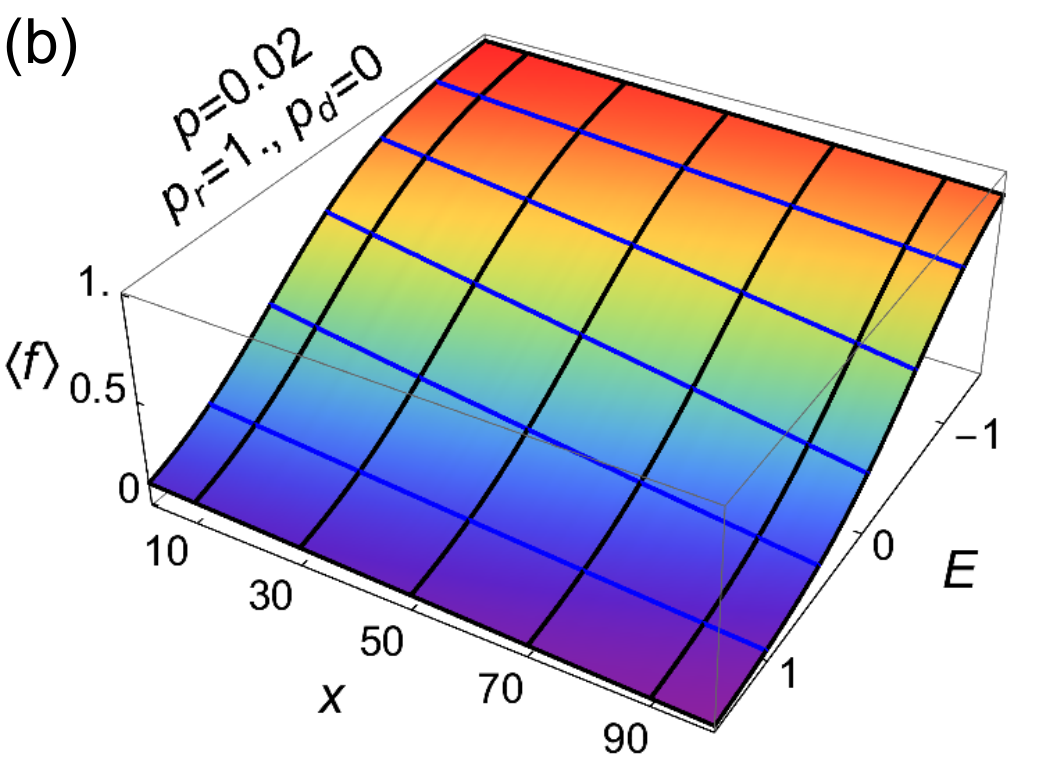}
\includegraphics[scale=0.36]{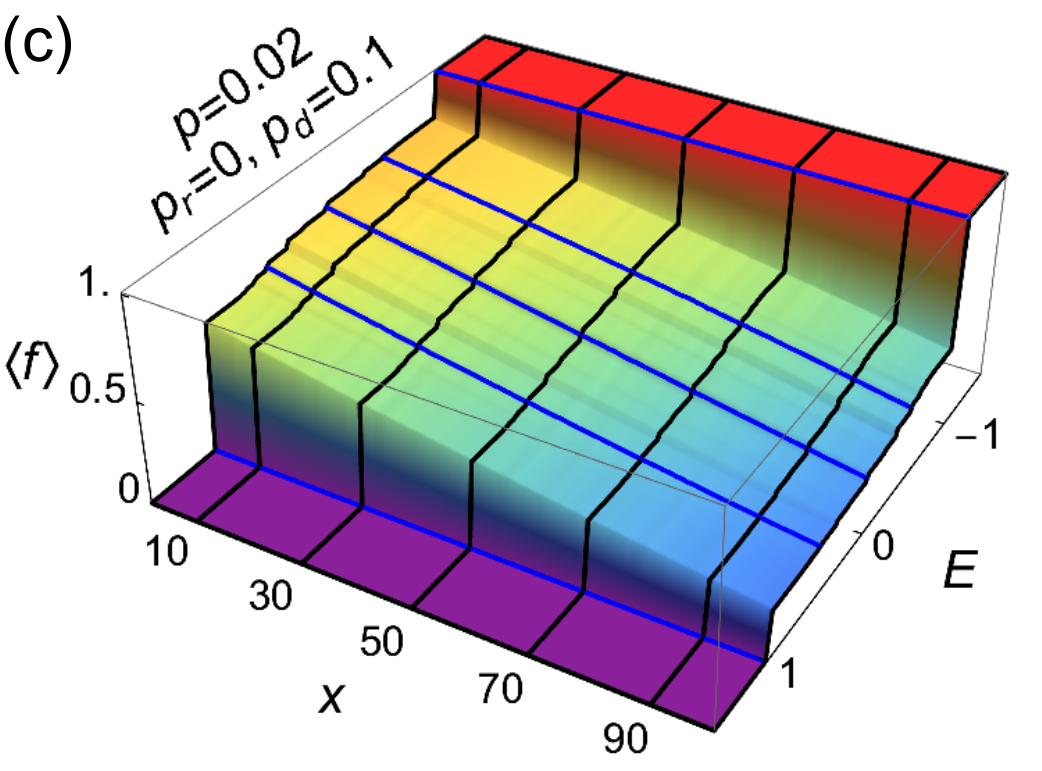}
\includegraphics[scale=0.36]{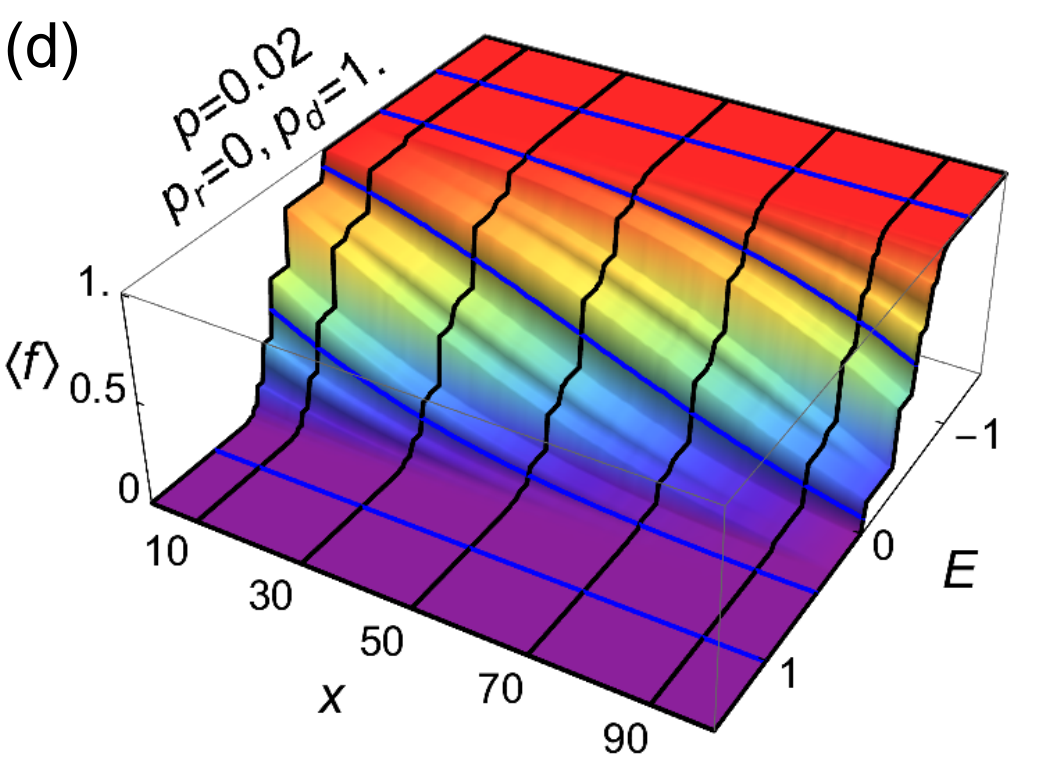}\\[3mm]
\includegraphics[scale=0.36]{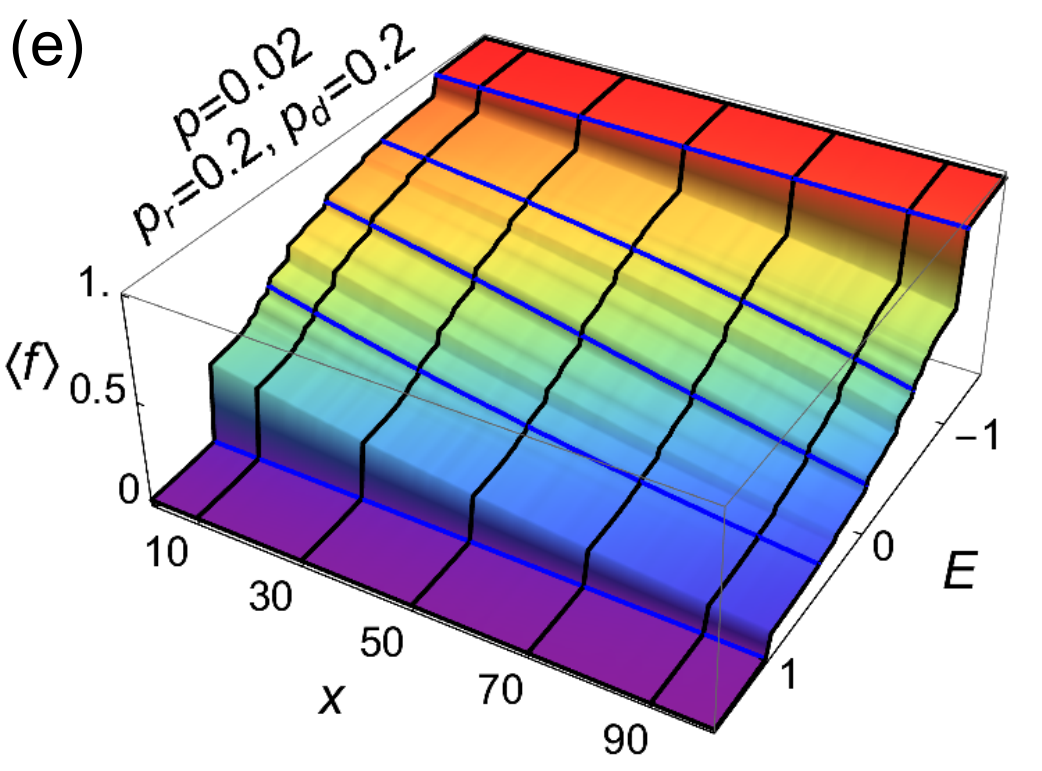}
\includegraphics[scale=0.36]{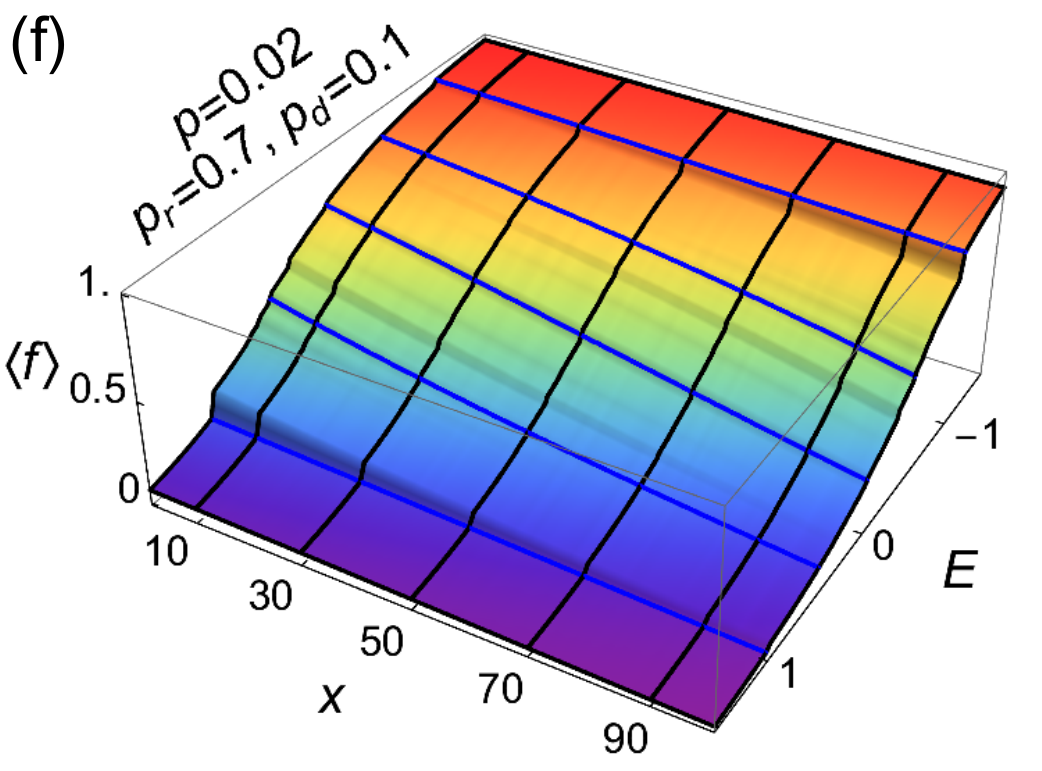}
\includegraphics[scale=0.36]{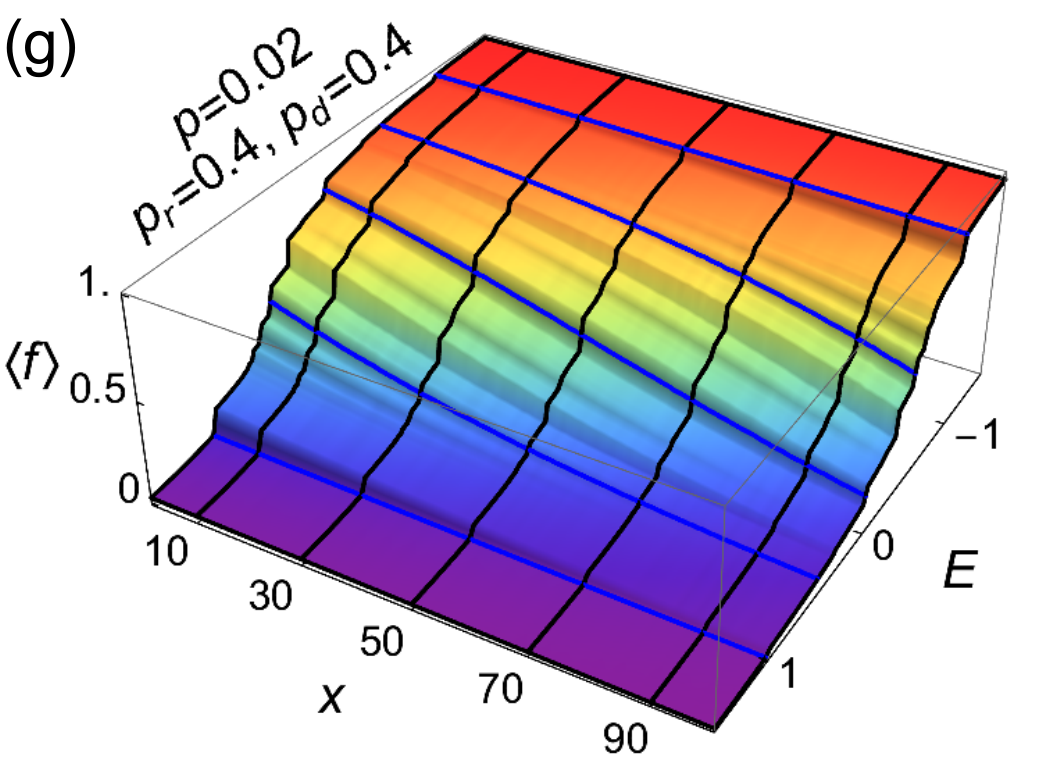}
\includegraphics[scale=0.36]{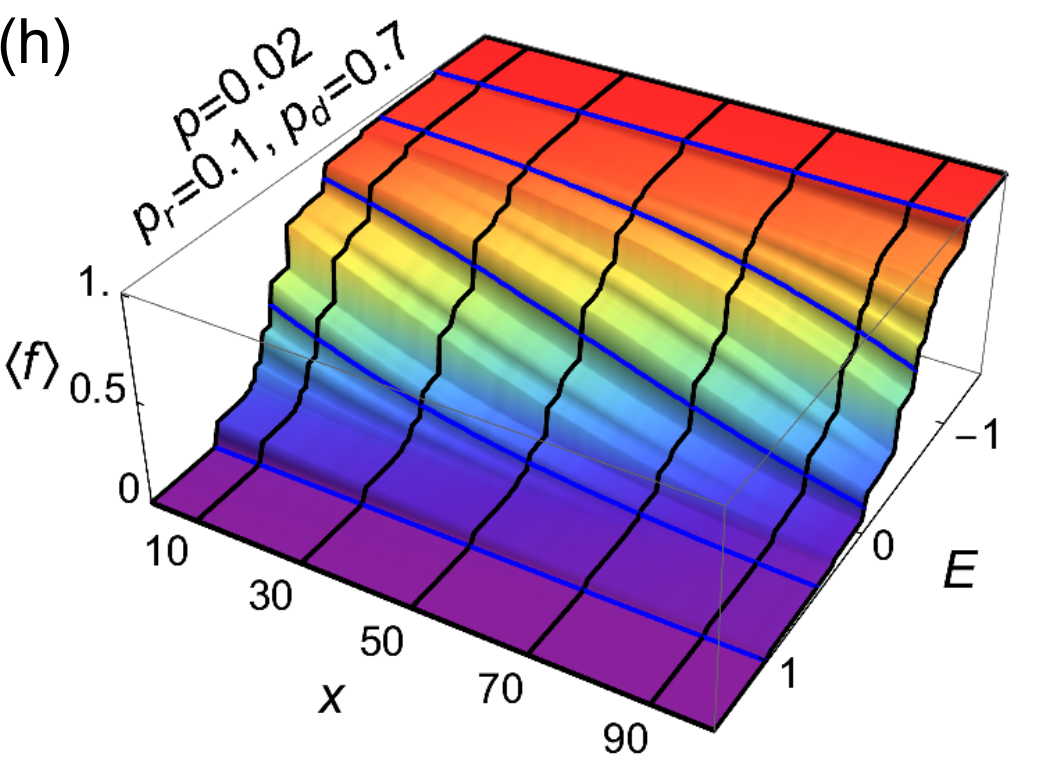}
\caption{Energy distribution functions $\av{f}$ under the effect of energy redistributing decoherence and
  energy dissipating decoherence. The parameters are the same as in \fig{4}, apart from the fact
  that the degree of decoherence $p=0.02$.}
\label{fig:9}
\end{figure}

\begin{figure}[htb]
\centering
\includegraphics[scale=0.7]{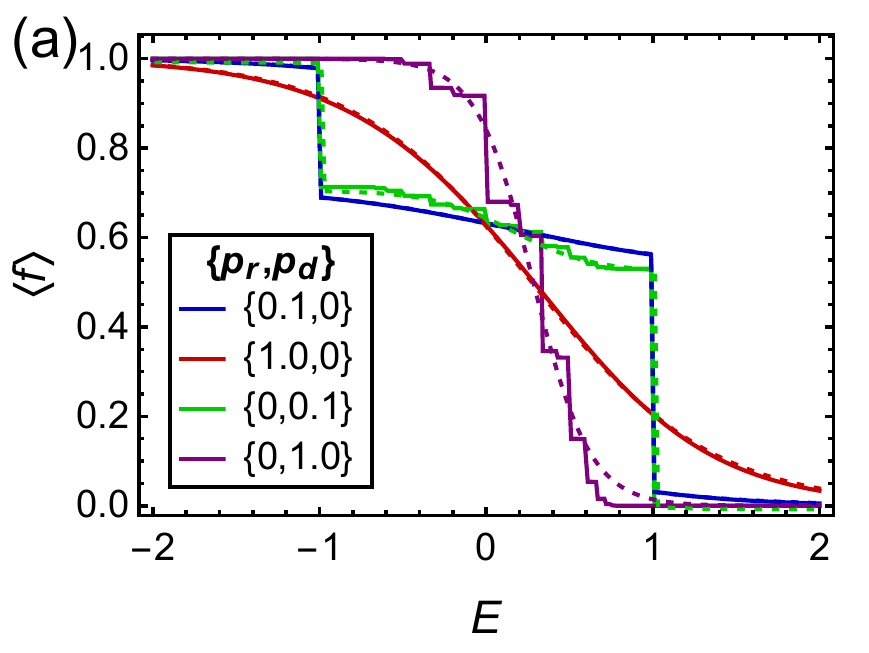}
\includegraphics[scale=0.7]{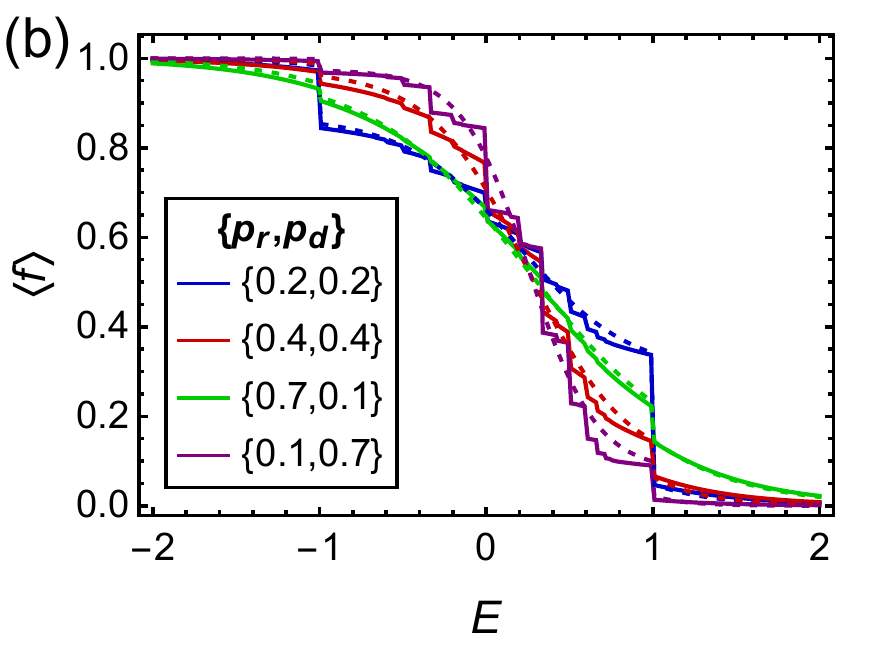}
\caption{Energy distribution function at $x=20$. In comparison to \fig{5} several smaller steps are
  observed. The model in \eq{model} cannot reproduce these steps. Apart from this it matches the
  numerical data.}
\label{fig:10}
\end{figure}

\Fig{9} shows the ensemble averaged energy distribution function in the case of weak decoherence
$p=0.02$. The shape of the distribution function agrees qualitatively with the results for $p=0.2$
shown in \fig{4}. However, if $p_d>0$ we observe, apart from the steps at $\mu_{S/D}= \pm1$, several
smaller steps. These steps can be seen even more clearly in \fig{10}, where $\av{f(x=20)}$ is
shown. They cannot be reproduced by the model \eq{model}, which apart from that fits very well to
the numerical data. The small steps in $\av{f}$ can be attributed to the steps in the
zero-temperature Fermi function of the energy dissipating decoherence reservoirs. For low $p$, a
decoherence configuration consists of only some few decoherence reservoirs (in average two for
$p=0.02$ and $N=100$), which are either energy conserving, energy redistributing or
dissipating. This restricts strongly the number of possible Fermi energies and causes jumps of
considerable height in $\av{f}$. This behavior becomes even more pronounced if $p_d \sim 1$, compare
the solid green and purple curves in \fig{10}~(a). If $p$ is not close to zero, the Fermi energy can
adopt many different values and hence, the steps in the distribution function vanish [\footnote{The
  distribution function will show many very small steps, which cannot be distinguished from a
  continous curve.}]. This effect is smoothed out if a nonzero bath temperature is assumed. For the
same reason this effect is not observed for energy redistributing decoherence, where the hot
electrons relax to Fermi function of nonzero temperature. Jumps in the distribution functions are
reported also in \cite{Apostolov2014}, where a chain with superconducting leads is modeled. In both
cases the jumps are due to a characteristic interaction. While in this work jumps are due to a
weakly distributed ($p$ low) but very effective ($p_d$ high) energy dissipating scattering, in
\cite{Apostolov2014} multiple Andreev reflections are present.

\section{Application to carbon nanotube experiments}
\label{sec:Appl}

\begin{figure}[htb]
\centering
\includegraphics[scale=0.75]{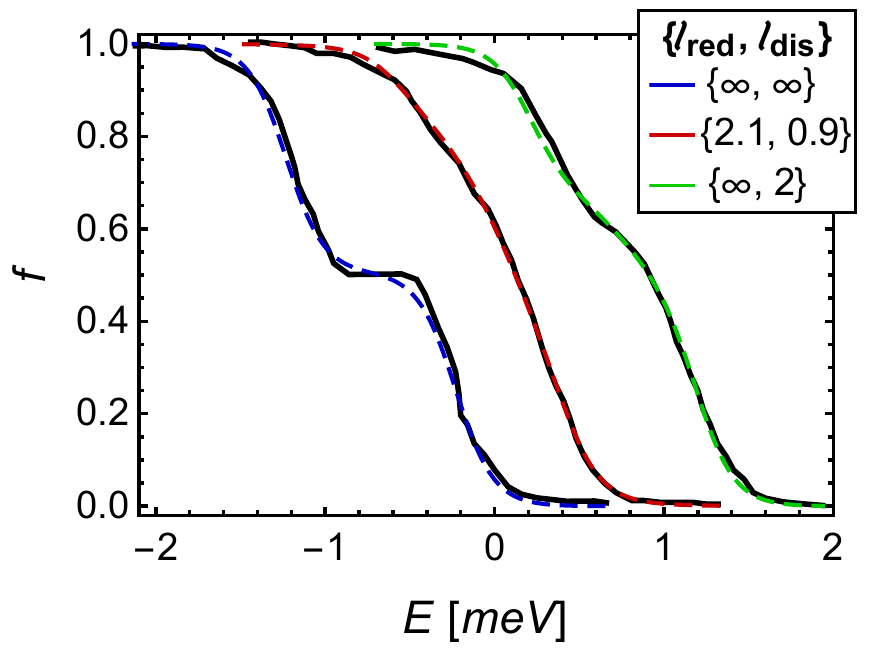}
\caption{Application of our model to the carbon-nanotube experiments in \cite{Chen2009}. The
  scattering lengths, given in the inset in units of the nanotube length, are determined by fitting
  our model to the data. Note that all curves have been shifted by some constant energy in order to
  show several measurements in a single figure.}
\label{fig:11}
\end{figure}

In this Section, we apply our model to the experiment in \cite{Chen2009}, where the energy
distribution function at a fixed position of a carbon nanotube is obtained from non-equilibrium
tunneling spectroscopy. A single-wall nanotube of length larger than $1 \un{\mu m}$ was studied at
temperature $\ta_{\text{bath}}= 1.3\un{K}$. The energy of the conduction electrons was controlled by
a gate voltage. The solid black curves in \fig{11} are the experimental data for a bias voltage of
$U= 1\un{mV}$ and the different gate voltages (from left to right)
$8. 660\un{V}, 8.285\un{V}, 8.070\un{V}$, taken from Figure~4~(a-c) of \cite{Chen2009}. Note that in
order to show the three curves more clearly in a single figure, we have shifted them by arbitrary
energy constants. At the large and the small gate voltage one observes (thermally broadened) double
step energy distribution functions consistent with the difference of $1\un{meV}$ between the source
and drain chemical potentials, which indicate ballistic transport in the carbon nanotube. For the
intermediate gate voltage, however, no remainder of the double step can be seen. As an explanation
it is suggested in \cite{Chen2009} that defect scattering is switched on by the intermediate gate
voltage, while it is largely suppressed by the other two values.

The experimental data agree well with results from our model, if the parameters are chosen
appropriately, see the dashed curves in \fig{11}. Instead of a microscopic tight-binding model, in
which all Carbon atoms of the nanotube are taken into account \cite{Reich2004}, we use a coarse
grained description: The nanotube of total length $L$ is divided into $Np=100$ segments of length
$\ell_{\phi}$ each. Therefore we represent each cell by a decoherence region of a one dimensional
chain. A fraction $p_d$ of these regions is energy dissipating, a fraction $p_r$ is energy
redistributing. These are the only fit parameters, because the bias voltage and the temperature are
set to the same values as in the experiment. The lengths $\ell_{\text{red/dis}}=1/(Npp_{r/d})$
measured in units of the nanotube length $L$ are given in the inset of \fig{11}. As
$\ell_{\text{red}} \geq \ell_{\text{dis}}$, we conclude that in this experimental setup it is more
likely that an electron of the Carbon nanotube exchanges energy with a heatbath than redistributing
it within the electron system. The heatbath could for example be the superconducting probe. The fact
that both scattering lengths are of the order of the nanotube length or larger shows that scattering
is really weak in this experiment.

\section{Conclusions}
\label{sec:Conclusions}

We have introduced a statistical model for the effects of decoherence, energy redistribution and
energy dissipation on the distribution function along a one-dimensional system driven out of
equilibrium by a source and a drain with different chemical potentials. Note that our model gives a
uniform Fermi-Dirac distribution throughout the system, if source and drain and the heat baths
coupled to the chain are in equilibrium. The essential idea of the model is to concentrate
decoherence exclusively in local regions, distributed stochastically in the system with probability
$p$. In these decoherence regions phase information about the electron state is lost entirely, while
its energy may be exchanged with a heat-bath (e.g. lattice vibrations) with probability $p_{d}$, or
be redistributed among the electrons with probability $p_{r}$, or otherwise remain
unchanged. One-dimensional quantum systems are divided by the decoherence regions into smaller
coherent subsystems, which makes our model computationally efficient. The probabilities, with which
the decoherence regions occur, are then inversely proportional to the scattering lengths of the
system. The energy distribution functions of the decoherence regions are coupled by rate equations,
which have been solved analytically for the case, that the coherent electron transmission between
decoherence regions can be described by $T(E)=1$. By ensemble averaging over the spatial decoherence
configurations, energy distribution functions are calculated everywhere in the quantum system, see
\fig{4} and \ref{fig:9}. Their shape depends on the distance from source and drain and on the
relative strength of the three types of decoherence (energy dissipating, redistributing, or
conserving). These non-equilibrium distribution functions turned out to be a weighted sum of three
Fermi functions, two of which belong to source or drain, respectively. Their contributions decay
exponentially with distance from source and drain, if energy dissipation or redistribution
occur. The third Fermi function belongs to the fraction of thermalized electrons. The temperature
and chemical potential depend on the distance from source and drain, and on the scattering
lengths. Our model also describes the electrical and thermal contact resistances at the electrodes,
which lead to discontinuities of the chemical potential and the temperature for weak decoherence.
That the model provides a useful tool for evaluating experiments, has been shown for data obtained
by non-equilibrium tunneling spectroscopy for a Carbon nanotube \cite{Chen2009}. The excellent
agreement of our model with the experimental data has allowed to determine the scattering lengths of
the nanotube. In the future, we will apply our model to cases, where $T(E) \neq 0$, for example to
DNA chains. We also plan to extend our model to pure dephasing by the possibility of adjusting
independently the degree of phase and momentum randomization \cite{Stegmann2012}. The extension of
our model to time dependent problems is also considered \cite{Rethfeld2002, Medvedev2010}.

\begin{ack}
  T.S. acknowledges financial support from CONACYT Proyecto Fronteras 952 ``Transporte en sistemas
  peque\~nos, cl\'asicos y cu\'anticos'' and from the PAPIIT-DGAPA-UNAM research grant IA101618
  ``Transporte electr\'onico en nano-estructuras de carbono''. We thank Francois Leyvraz for helpful
  discussions.
\end{ack}

\section*{References}
\bibliographystyle{iopart-num}
\bibliography{./enrel}

\end{document}